\documentclass[twocolumn,preprintnumbers,amsmath,amssymb,aps, prb,floatfix,groupedaddress]{revtex4-2}
\usepackage[]{graphics} 
\usepackage{bm}       
\usepackage{epsfig}
\usepackage{braket}
\usepackage{comment}
\usepackage{multirow}
\usepackage{tikz}
\usepackage{color}
\usepackage{pgffor}
\usepackage{siunitx}
\usepackage[section]{placeins} 
\usepackage[caption=false]{subfig}

\usepackage{newunicodechar}
\newunicodechar{−}{\textminus}

\begin{document}
\title[Article Title]{Correlation tuned Fermi-arc topology in a Weyl ferromagnet }


\author{Yiran Peng$^{1}$, Rui Liu$^{1}$, Pengyu Zheng$^{1}$}
\author{Zhiping Yin$^{1,2}$}
\email{yinzhiping@bnu.edu.cn}



\affiliation{ $^{1}$School of Physics $\&$ Astronomy and Center for Advanced Quantum Studies, Beijing Normal University, Beijing 100875, China}
\affiliation{ $^{2}$Key Laboratory of Multiscale Spin Physics (Ministry of Education), Beijing Normal University, Beijing 100875, China}

\date{\today}

\begin{abstract}
Electrons on Fermi arcs (FAs), a hallmark of Weyl semimetals, exhibit chiral transport harboring chiral anomaly, negative magnetoresistance, and Majorana zero modes. While FAs were observed in exemplary Weyl semimetal TaAs and Co$_3$Sn$_2$S$_2$, the manipulation of FAs has been rarely explored. Here we take Co$_3$Sn$_2$S$_2$ as an example and demonstrate that tuning the electronic correlation strength is an effective way to control the topology and connectivity of FAs.  After achieving a good agreement with experimentally measured band structure by employing combined density functional theory and dynamical mean field theory (DFT+DMFT) calculations, we show that the experimental charge dynamics are well reproduced by DFT+DMFT calculations but not DFT calculations. Electronic correlation renormalizes the bands around the Fermi level and modifies the energy and location of Weyl points, and the resulting FAs. In particular, on the Co-terminated surface, the FAs are formed by connecting Weyl points located in adjacent Brillouin zones in DFT+DMFT calculations and experiments, in strong contrast to the FAs connecting Weyl points within the same Brillouin zone in DFT calculations. We further show the evolution of FAs with correlation and reveal a topological change of the FAs on the Sn-terminated surface at stronger correlation strength. Our study sheds new light on experimental manipulation of FAs to improve the electronic properties of correlated Weyl semimetals. 
\end{abstract}

\maketitle

\section{Introduction}
In the past decade, Weyl semimetals (WSMs) have become a significant research focus in the field of topological matter due to their remarkable features including extremely large linear magnetoresistance\cite{PhysRevB.94.235154,liu2018giant}, ultrahigh mobility\cite{Nagpal_2020}, chiral anomaly effect\cite{PhysRevX.5.031023,zhang2016signatures,PhysRevB.93.121112,niemann2017chiral}, gravitational anomaly effect\cite{gooth2017experimental}, and strong intrinsic anomalous Hall and spin Hall effects\cite{PhysRevLett.117.146403,PhysRevLett.107.186806,PhysRevLett.107.127205}. 
The nontrivial topological properties of these materials originate from the presence of band-touching points known as Weyl points with opposite chirality in the bulk electronic  band structures. These Weyl points can arise when either inversion symmetry (IS)\cite{PhysRevX.5.011029,huang2015weyl,science.aaa9297,PhysRevX.5.031013,yang2015weyl} or time-reversal symmetry (TRS)\cite{PhysRevB.83.205101,RevModPhys.90.015001,PhysRevLett.107.186806,PhysRevLett.117.236401} is broken. 
These Weyl points possess definite chirality and give rise to a distinct surface state called ``Fermi arc", which is an open curve connecting two Weyl points with opposite chirality on the surface Brillouin zone of the material\cite{PhysRevB.83.205101,PhysRevX.5.011029,huang2015weyl,science.aaa9297,PhysRevX.5.031013,yang2015weyl}. 
While the presence of Weyl points ensures the emergence of Fermi arcs, the characteristics of the Fermi arcs, such as the shape of the curve and the connectivity between Weyl points, are sensitive to the band structure, especially the energies and locations of these Weyl points in the material\cite{PhysRevB.92.115428,PhysRevB.93.161112}. 
The profile of the Fermi arcs and the connectivity of the Weyl points exert a direct influence on the magnetoelectric dynamics, including quantum oscillations, and transport properties of the electrons that encompass both bulk and surface conduction\cite{potter2014quantum,moll2016transport,science.aav2334}. 
However, there has not been much discussions on what affects the profile and the connectivity of the Fermi arcs in the literature.

\begin{figure}[ht]
	\includegraphics[width=\columnwidth]{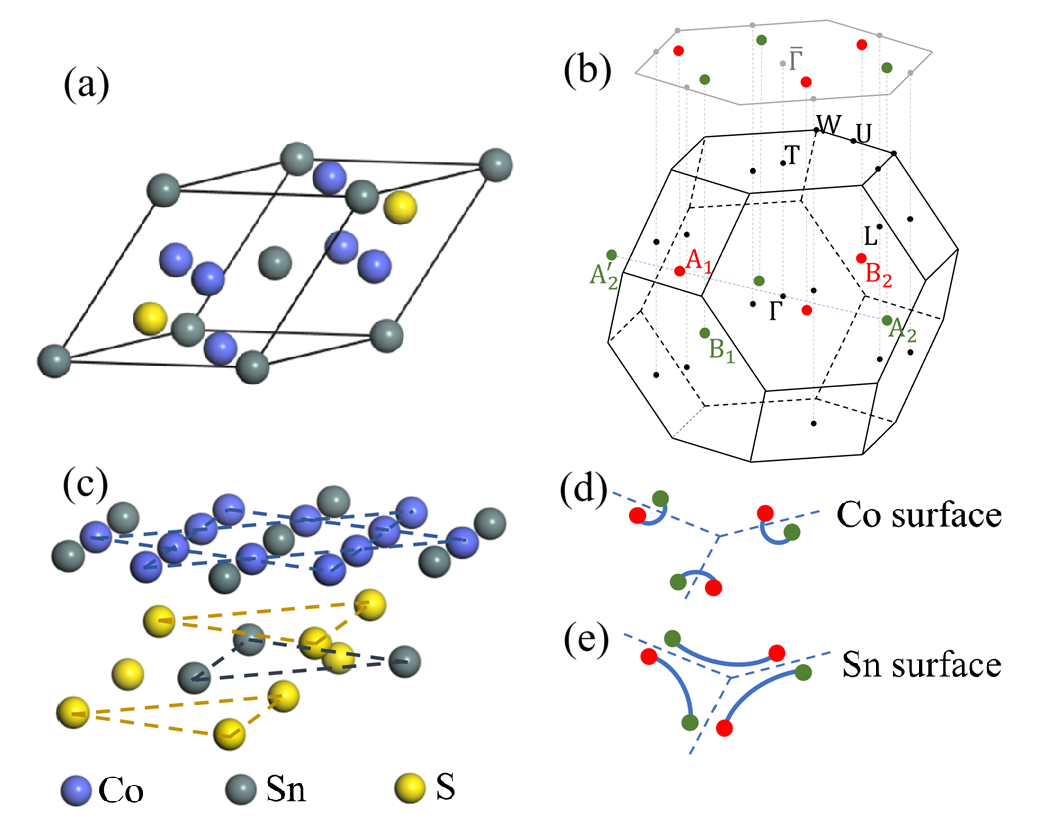}
	\caption{
		(a) and (c): Crystal structure of Co$_3$Sn$_2$S$_2$; (b) The bulk Brillouin zone contains three pairs of Weyl points and their projections on the (001) surface, with red and green dots representing positive and negative chiralities of Weyl points, respectively. (d) and (e) The Fermi arc topology on Co and Sn surfaces, respectively\cite{science.aav2334}. }	
	\label{struct}
\end{figure}

The ferromagnetic (FM) Kagome metal Co$_3$Sn$_2$S$_2$ has a Curie temperature of 177 K and is composed of Sn, Co$_3$Sn, and S layers. Within the Co$_3$Sn layer, the Co atoms form a quasi-two-dimensional Kagome lattice with Sn atoms located at the center of each hexagon (shown in Fig.~\ref{struct}(c))\cite{VAQUEIRO2009513,xu2020electronic,zaac.200400561}. 
Experimental evidence confirmed its chiral anomaly\cite{liu2018giant,yoshikawa2022non}, significant anomalous Hall effect\cite{liu2018giant,ikeda2021critical,wang2018large,okamura2020giant} and anomalous Nernst effect\cite{PhysRevX.9.041061,adma.201806622,PhysRevMaterials.4.024202}, linear bulk band dispersion around a Weyl point, as well as Fermi arcs on different surfaces\cite{science.aav2873,ikeda2021two,howard2021evidence}. 
Consequently, Co$_3$Sn$_2$S$_2$ has been further identified as a TRS-breaking WSM containing three pairs of Weyl points in the first Brillouin zone (BZ) (shown in Fig.~\ref{struct}(b))\cite{liu2018giant,PhysRevB.97.235416,bernevig2022progress,science.aav2873}. 
It is noteworthy that distinct Fermi arc profiles and connectivities of Weyl points were observed on three different surface termination of Co$_3$Sn$_2$S$_2$ using scanning tunneling microscopy (STM)(shown in Fig.~\ref{struct}(d, e))\cite{science.aav2334}. 
In FM ground state, Co$_3$Sn$_2$S$_2$ exhibits a moderate level of electron correlation strength\cite{xu2020electronic}. 
The flat band near the Fermi level is influenced by both electron correlation effects and magnetization\cite{xu2020electronic,nag2022correlation}. 
Many studies have demonstrated that the presence of sufficient electron correlation strength can induce the opening of an energy gap at the Weyl points, thereby disrupting the Weyl states\cite{PhysRevLett.109.196403,PhysRevLett.109.066401,PhysRevB.87.161107,PhysRevLett.113.136402,JPSJ.83.094710,PhysRevB.92.241109,PhysRevLett.114.237001,morimoto2016weyl,PhysRevB.94.075115,PhysRevB.94.241102,PhysRevB.95.201102}. 
Furthermore, experimental evidence confirms the significance of spin-orbit coupling (SOC) in shaping the formation of a Weyl semimetal in Co$_3$Sn$_2$S$_2$\cite{liu2022direct}. 
Considering its ferromagnetism, correlation strength, and SOC effects, Co$_3$Sn$_2$S$_2$ serves as an excellent platform for investigating the factors influencing the topological surface states, i.e., Fermi arcs, in WSMs.

To take into account the electronic correlation effect, we employ density functional theory combined with dynamical mean field theory (DFT+DMFT) to compute the electronic structure of Co$_3$Sn$_2$S$_2$ and compare with results from DFT calculations. While the DFT calculated band structure shows clear discrepancy with the angle-resolved photoemission spectroscopy (ARPES) measured band structure\cite{science.aav2873,PhysRevB.104.205140}, the DFT+DMFT calculations reproduce very well the ARPES experimental results with a Hubbard U=4.0 eV and Hund's coupling J=0.8 eV. The same set of U and J values are used in DFT+DMFT to compute the charge dynamics, topological characteristics and the Fermi arcs on different surfaces. 
The experimental in-plane optical conductivity in both the paramagnetic and ferromagnetic states\cite{PhysRevLett.124.077403} are well reproduced by DFT+DMFT calculations but not DFT calculations. The spin polarization in the ferromagnetic state gives rise to two additional peak structure in the low frequency region where the first peak is found to originate from a combined effect of spin polarization, spin-orbit coupling and electronic correlation. 

These good agreement between experiments and DFT+DMFT calculations allow us to compute reliably the topological properties including the energies and locations of the Weyl points, the connectivity of the Weyl points, i.e., Fermi arcs, and the profile (or shape) of the Fermi arcs on different (Co-, Sn-, and S-terminated) surfaces. In the DFT calculations, the Fermi arcs on both the Co- and Sn-terminated surfaces connect Weyl points of opposite chirality within the same Brillouin zone (BZ), while the Fermi arcs on the S-terminated surface almost merge into bulk states. However, DFT+DMFT calculations reveal distinct profile and connectivity of the Fermi arcs: the Fermi arcs on the Co-terminated surface connect Weyl points across two adjacent BZs, while those on the Sn-terminated surface remain connecting Weyl points within the same BZ but have very different profile. The DFT+DMFT calculated Fermi arcs agree well with the scanning tunneling microscopy experiments\cite{science.aav2334} which implies that electronic correlation has a strong impact on the profile and connectivity of the Fermi arcs in correlated semimetals. 

To demonstrate how the Fermi arcs evolve with the strength of electronic correlation, we build a model Hamiltonian mixing the DFT+DMFT Hamiltonian $H_{DFT+DMFT}$ and DFT Hamiltonian $H_{DFT}$, i.e., $H(x)=x* H_{DFT+DMFT} + (1-x) * H_{DFT}$, where $x=1$ corresponds to DFT+DMFT calculations and $x=0$ corresponds to DFT calculations. We find that the Fermi arcs on the Co-terminated surface undergo a topological transition (i.e., a change of connectivity of the Weyl points) around $x=0.95$ whereas the Fermi arcs on the Sn-terminated surface undergo a topological transition around $x=1.03$. To verify the latter topological transition, we carry out DFT+DMFT calculations with larger values of U=6.0 eV and J=0.9 eV and confirm that indeed the Fermi arcs on the Sn-terminated surface are now connecting Weyl points across two adjacent BZs. Our work highlights the important role of electron correlation in determining the topology and connectivity of Fermi arcs in correlated Weyl semimetals which may be used as a strategy for experimental manipulation of Fermi arcs. \\

\section{Results}

\subsection{Electronic structures and topological characters}



\begin{figure}[ht]
	\includegraphics[width=\columnwidth]{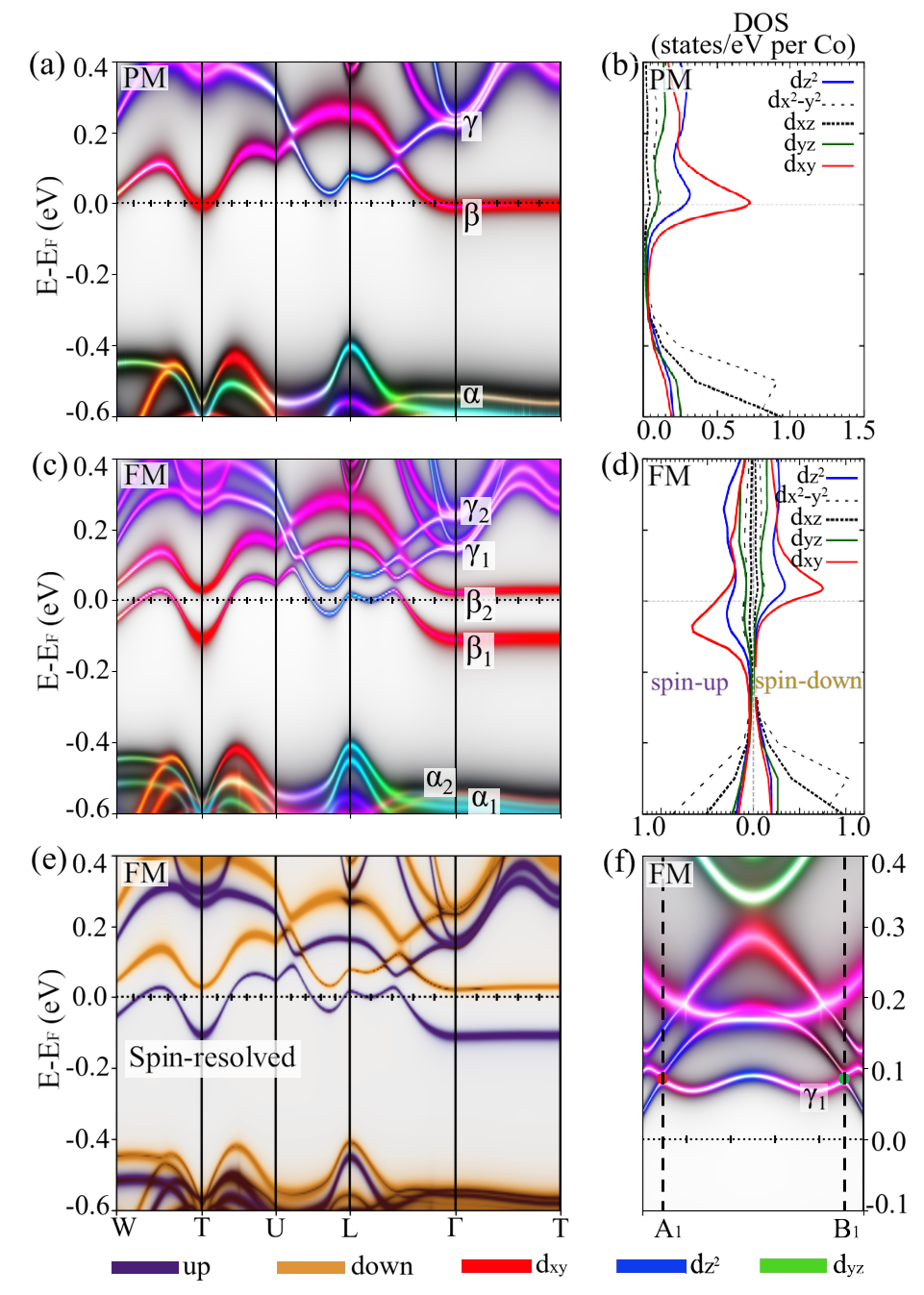}
	\caption{
		Momentum, energy, and orbital/spin-resolved spectra and projected density of states (DOS) of Co$_3$Sn$_2$S$_2$ in the PM and FM states calculated by DFT+DMFT with SOC. (a) and (c): The orbital-resolved spectra in PM and FM states, respectively. The red, green and blue colors denote Co $3d_{xy}$, Co $3d_{yz}$ and Co $3d_{z^2}$ orbitals, respectively. (b) and (d): The DOS in PM and FM state, respectively. (e): The spin-resolved spectra in FM state. The violet and orange colors denote the spin up and spin down orbitals, respectively.   (f): The orbital-resolved spectra in FM state along the $k$-path crossing a pair of Weyl points. The red and green points represent Weyl points with positive and negative chiralities, respectively. 
	}	
	\label{fatband}
\end{figure}

Since both electronic correlation and SOC play important roles, we first show in Fig.~\ref{fatband} the 
momentum, energy, and orbital/spin-resolved spectra and projected density of states (DOS) of Co$_3$Sn$_2$S$_2$ in the paramagnetic (PM) and FM states calculated by DFT+DMFT with SOC. We note that the momentum and energy-resolved spectra calculated by DFT+DMFT agrees well with experimental ARPES results along the measured $k$-paths.\cite{science.aav2873,PhysRevB.104.205140,SI}

In the PM state (Fig.~\ref{fatband}(a)), the lowest conduction band (denoted as $\beta$ band) is dominated by the Co $3d_{xy}$ orbital while the next conduction band  (denoted as $\gamma$ band) is dominated by the Co $3d_{z^2}$ orbital with some mixture of the Co $3d_{yz}$ orbital. Along the $\Gamma$-$T$ path, the $\beta$ band is quite flat and lies at the Fermi level, which gives rise to a large peak centered at the Fermi level in the projected DOS of Co $3d_{xy}$ orbital (Fig.~\ref{fatband}(b)). 

Entering the FM state, the spin-exchange interaction splits the conduction $\beta$ ($\gamma$) band into a spin-up $\beta_{1}$ ($\gamma_1$) band and spin-down $\beta_{2}$ ($\gamma_2$) band (Fig.~\ref{fatband}(c,d,e)). The magnitude of the spin-exchange varies from $\sim$60 meV to $\sim$130 meV at different momentums. It is noted that, the $\beta_{2}$ and $\gamma_2$ bands comprise a certain proportion of spin-up components due to SOC (Fig.~\ref{fatband}(e)), which has important implications for the charge dynamics in the low frequency region as discussed in the next section.

Along the $U$-$L$-$\Gamma$ path, a band inversion occurs between the $\beta$ band and the $\gamma$ band in the PM state. 
SOC induces a cross-avoiding gap between the $\beta$ and $\gamma$ bands, through which we can define a Fermi curve\cite{PhysRevB.76.045302}  
This Fermi curve represents the momentum-dependent ``Fermi energy" and separates the hypothetically occupied states below this curve and unoccupied states above this curve. 
By characterizing all the states below the Fermi curve, the Fu-Kane parity criterion on topological insulators can be used to characterize the topological properties of metals. 
The $Z_2$ topological invariant for all the bands below this Fermi curve is calculated to be $(1;000)$, indicating a strong $Z_2$ topological index which gives rise to the Dirac-cone topological insulator (TI) states on the (001) surface (see supplementary materials). 
In the FM state, without considering SOC, the $\beta_{1}$ and $\gamma_{1}$ band would form a nodal ring around the $L$ point.  However, SOC gaps out the nodal ring except one pair of Weyl points with opposite topological charges ($+1$ and $-1$ Chern numbers) are preserved as shown in  Fig.~\ref{fatband}(f). \\

\subsection{Charge dynamics}

To further validate the accuracy of the DFT+DMFT calculations, we use both DFT and DFT+DMFT to calculate the optical conductivity of Co$_3$Sn$_2$S$_2$ in both FM and PM states and compare them with experimental results.

\begin{figure}[!ht]
	\includegraphics[width=1.02\columnwidth]{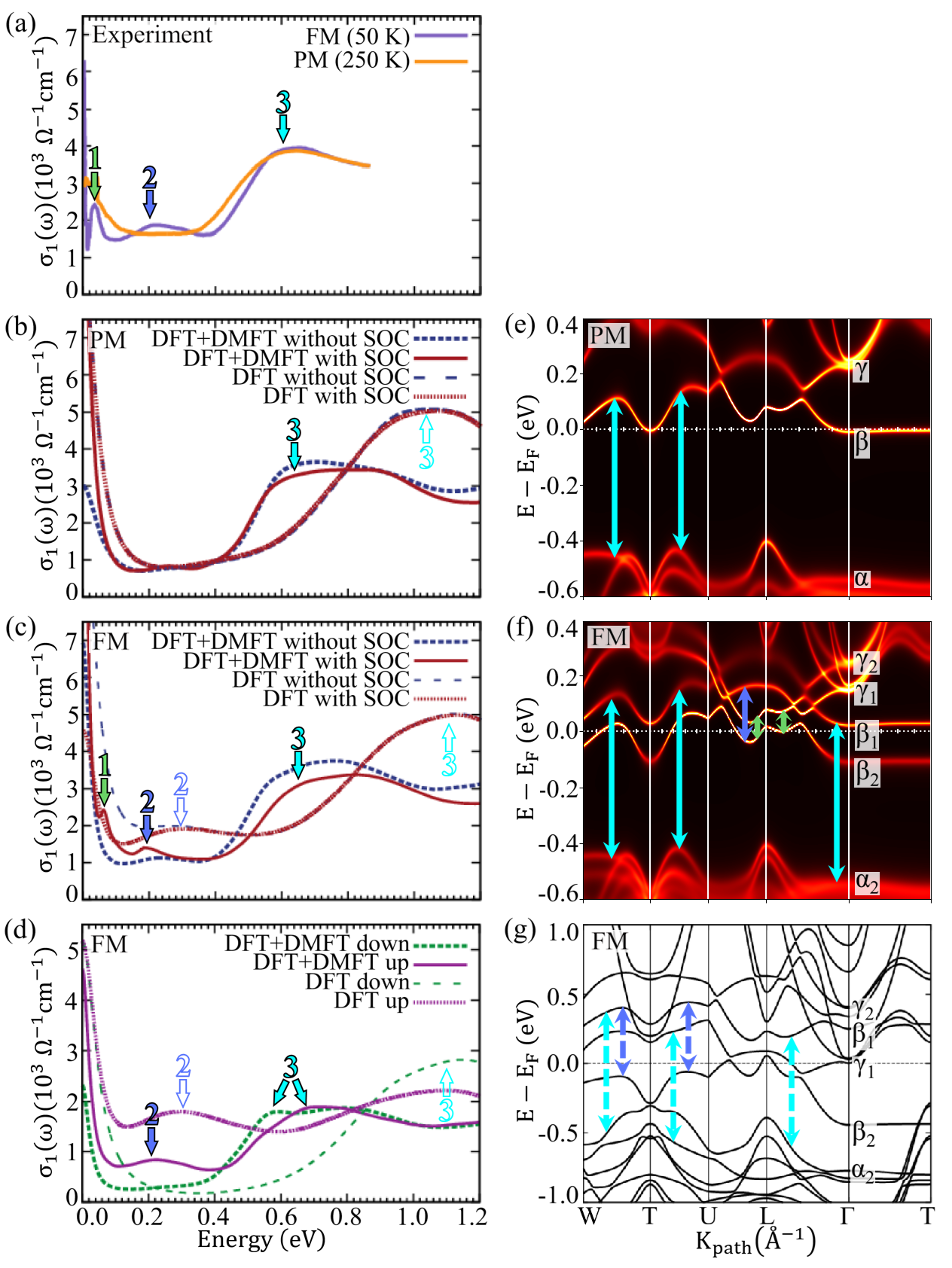}
	\caption{
		The optical conductivity  $\sigma_{1}(\omega)$ and band structure of Co$_3$Sn$_2$S$_2$ in the PM and FM states. Three interband absorption peaks from low to high energy are denoted by green, blue and cyan arrows. (a) The experimental optical conductivity $\sigma_{1}(\omega)$ up to 0.9 eV at different temperatures reported in Ref.~\cite{PhysRevLett.124.077403}. (b,c) The DFT and DFT+DMFT calculated $\sigma_{1}(\omega)$ in the PM state (b) and FM state (c). (d)The DFT and DFT+DMFT calculated spin-resolved $\sigma_{1}(\omega)$ in the FM state without SOC. (e,f) The band structures of DFT+DMFT calculation with SOC in PM and FM states , respectively. (g) The band structures of DFT calculation with SOC in FM state
	}
	\label{opc}
\end{figure}

The experimentally measured real part of the in-plane optical conductivity $\sigma_{1}(\omega)$ of Co$_3$Sn$_2$S$_2$ is depicted in Fig.~\ref{opc}(a) Ref.~\cite{PhysRevLett.124.077403}, which is consistent with another experimental data reported in Ref.~\cite{xu2020electronic}. 
In the PM state (T=250 K), apart from the Drude peak originating from the intraband response at zero frequency, a Lorentz peak originating from the interband transition appears at approximately 0.62 eV (peak 3, indicated by the cyan arrow in Fig.~\ref{opc}(a))\cite{PhysRevLett.124.077403}. 
In FM state (T=50 K), the optical conductivity exhibits a zero-frequency Drude peak and three Lorentz peaks. These peaks are located at 0.038 eV (Peak 1, indicated by the green arrow in Fig.~\ref{opc}(a)), 0.2 eV (Peak 2, blue arrow in Fig.~\ref{opc}(a)), and 0.65 eV $cm^{-1}$ (Peak 3, cyan arrow in Fig.~\ref{opc}(a))
In the DFT calculation of the non-magnetic state, a single Lorentz peak (peak 3, cyan open symbols) is observed at approximately 1.0 eV which has an obvious difference with experiments.
After considering the correlation effect, peak 3 of the PM optical conductivity (Fig.~\ref{opc}(b)) shifts to approximately 0.65 eV in the DFT+DMFT calculation. 
This indicates that electronic correlation effect renormalizes the overall DFT band structure by a factor of 1.5, which is consistent with the mass enhancement estimated from the self-energy of the Co $3d$ orbitals (1.5 for $3d_{xy}$ and $3d_{z^2}$ orbitals, 1.3 for the rest). 

When SOC is not considered, the optical conductivity in the FM state exhibits two characteristic peaks: peak 2 at approximately 0.2 eV (0.3 eV) and peak 3 around 0.62 eV (1.1 eV) in the DFT+DMFT (DFT) calculations(Fig.~\ref{opc}(c,d)). It is noted that the DFT calculations are consistent with previous reports\cite{xu2020electronic, PhysRevLett.124.077403}.
With SOC is taken into account, peak 2 and peak 3 undergo some subtle changes in the DFT+DMFT and DFT calculations compared to the calculations without SOC.
Surprisingly, a distinct peak 1 emerges near 0.060 eV in the DFT+DMFT calculations, closely resembling experimental findings at low temperatures\cite{xu2020electronic, PhysRevLett.124.077403}. 
In strong contrast, no new peak appears below peak 2 in the DFT calculations.
This suggests that SOC, ferromagnetism and electronic correlation work collaboratively to give rise to the emergence of peak 1 at low temperature.
We note that our DFT+DMFT calculation explains naturally the origin of peak 1 whereas a previous DFT+DMFT calculation reported three different peaks in the optical conductivity near peak 1 \cite{xu2020electronic}.

We further identify the vertical transitions which make dominating contributions to the three peaks in the optical conducitity in Fig.~\ref{opc}(e,f,g). 
Combining the fatband analysis and optical conductivity obtained by DFT+DMFT calculations, we find that peak 1 mainly originates from transitions between the $\gamma_{1}$ and $\gamma_{2}$ bands along the $U$-$L$-$\Gamma$ path. 
After analyzing the band structure before and after magnetization, we propose that peak 2 in the optical conductivity mainly originates from the transition between the $\beta_{1}$ and $\gamma_{1}$ bands along the $U$-$L$ path. 
Peak 3 primarily arises from the transition between the $\alpha_{2}$ and $\beta_{2}$ band. 
For peaks 3, the outcomes obtained from DFT calculations are similar to those from DFT+DMFT analysis. 
Meanwhile, peak 2 in the DFT calculations is mainly caused by the transition between the $\beta_{1}$ band and $\gamma_{1}$ band along the $W$-$T$-$U$ path.

Our findings demonstrate excellent agreement between the DFT+DMFT calculated and experimental optical conductivity in the PM and FM states, which further confirms the accuracy of the DFT+DMFT calculation. Therefore, it is expected that the DFT+DMFT calculations can accurately predict the topology of the band structure and the surface states, in particular, the energies and locations of the Weyl points, and the topology and connectivity of the Fermi arcs. \\

\subsection{Fermi arcs}

Fermi arcs play a pivotal role in elucidating the non-trivial topological properties of Co$_3$Sn$_2$S$_2$.
A Fermi arc connects two Weyl points of opposite chirality in the bulk. However, there is a lack of comprehensive discussions regarding the factors that govern the connecting pattern of the Weyl points and the topology of the resulting Fermi arcs.
In the FM state of Co$_3$Sn$_2$S$_2$, the $\beta_{1}$ band and $\gamma_{1}$ band are renormalized by electronic correlation and shifted by spin-exchange interaction. Therefore, the energies and locations of the Weyl points formed by the crossing of the $\beta_{1}$ and $\gamma_{1}$ bands depend sensitively on the magnetization magnitude and electronic correlation strength. Compared to DFT calculations, the $\beta_{1}$ and $\gamma_{1}$ bands, as well as the Weyl points, are shifted upward in the DFT+DMFT calculations (Fig.~\ref{opc}(f,g)). 
Given the intrinsic connection between Fermi arcs and the topology of the bulk electronic structure \textit{via} a topological boundary mapping\cite{Mathai_2017,jia2016weyl}, 
it is interesting to explore how the Fermi arcs evolve with magnetization and electronic correlations in Co$_3$Sn$_2$S$_2$.

\begin{figure}[!ht]
	\includegraphics[width=\columnwidth]{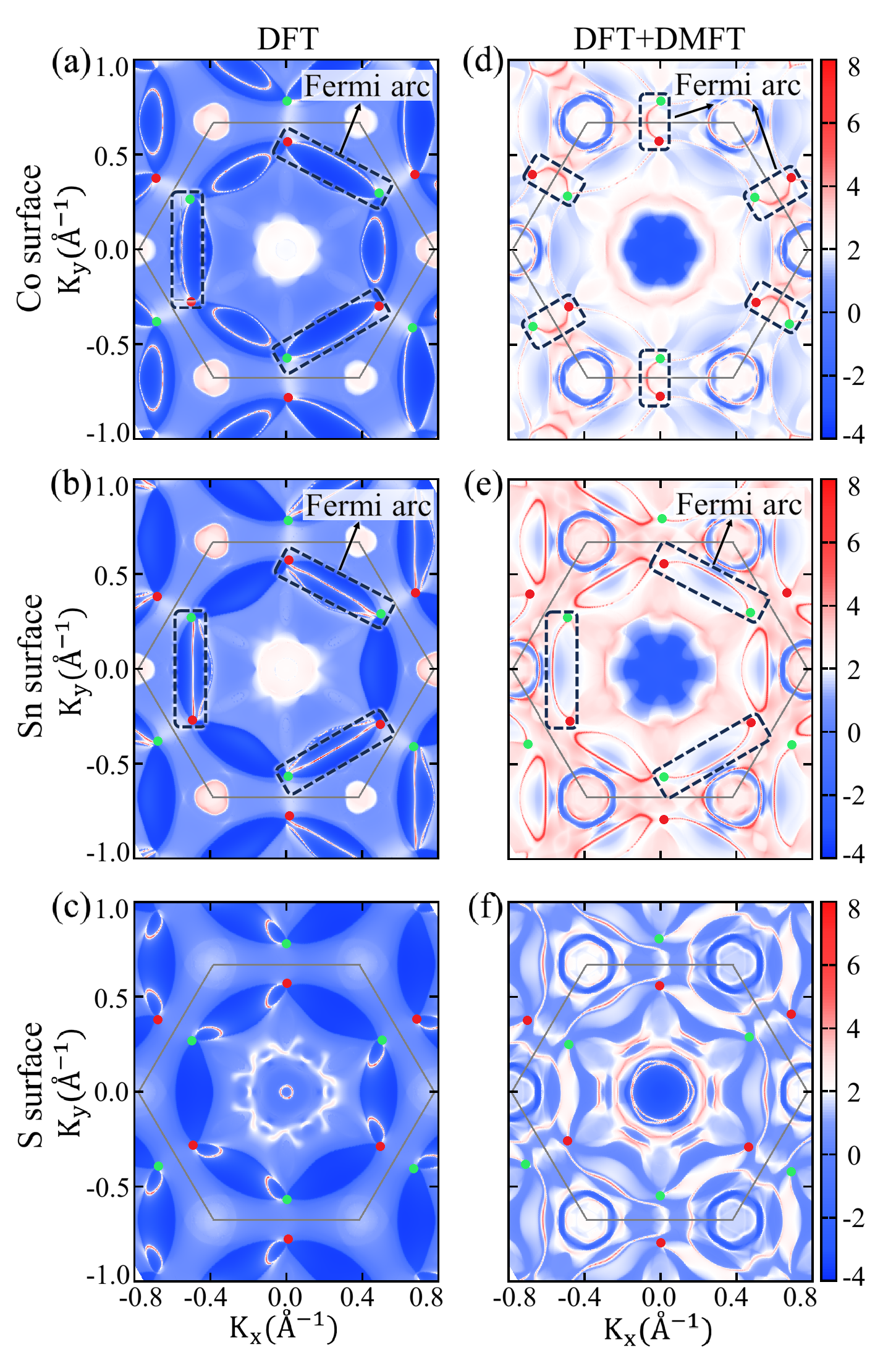}
	\caption{
		The Fermi arcs on the (001) surface for different terminations of Co$_3$Sn$_2$S$_2$ in the FM state. The DFT (a-c) and DFT+DMFT (d-f) calculated (001) surface states on the Co, Sn and S terminated surfaces, respectively. The pink and black points represent Weyl points with positive and negative chiralities, respectively. The dashed black box indicates the Fermi arcs connecting the Weyl points with opposite chirality. The dashed red box indicates the tendency of the Fermi arcs connecting the Weyl points between the adjacent Brillouin zones.  }	
	\label{sf1}
\end{figure}

To this end, we compute the Fermi arcs on different surfaces in the FM state using DFT and DFT+DMFT and show them in Fig.~\ref{sf1}.
In the DFT calculations (Fig.~\ref{sf1}(a,b,c)), it is evident that on both Co and Sn-terminated surfaces, Fermi arcs predominantly connect Weyl points of opposite chirality within the same BZ, whereas the Fermi arcs on the S-terminated surface merge into bulk states. 
On the other hand, in the DFT+DMFT calculations, the Fermi arcs on the Co-terminated surface connect Weyl points of opposite chirality between adjacent BZs (Fig.~\ref{sf1}(d)), in strong contrast to DFT calculations. 
While the Fermi arcs on the Sn-terminated surface also connect Weyl points of opposite chirality within a single BZ in the DFT+DMFT calculations(Fig.~\ref{sf1}(e)), they have some different topology from the DFT calculations(Fig.~\ref{sf1}(b)). The Fermi arcs on the S-terminated surface also merge into bulk states in the DFT+DMFT calculations (Fig.~\ref{sf1}(d)).
Note that, the DFT+DMFT calculated Fermi arcs on the Co and Sn-terminated surfaces (Fig.~\ref{sf1}(d,e)) aligns well with experimental findings\cite{science.aav2334} (reproduced in Fig.~\ref{struct}(d,e)), which demonstrates the importance of achieving accurate electronic structure for calculating reliably the topological properties of materials including Fermi arcs.

\begin{figure*}[!ht]
	\includegraphics[width=2.0\columnwidth]{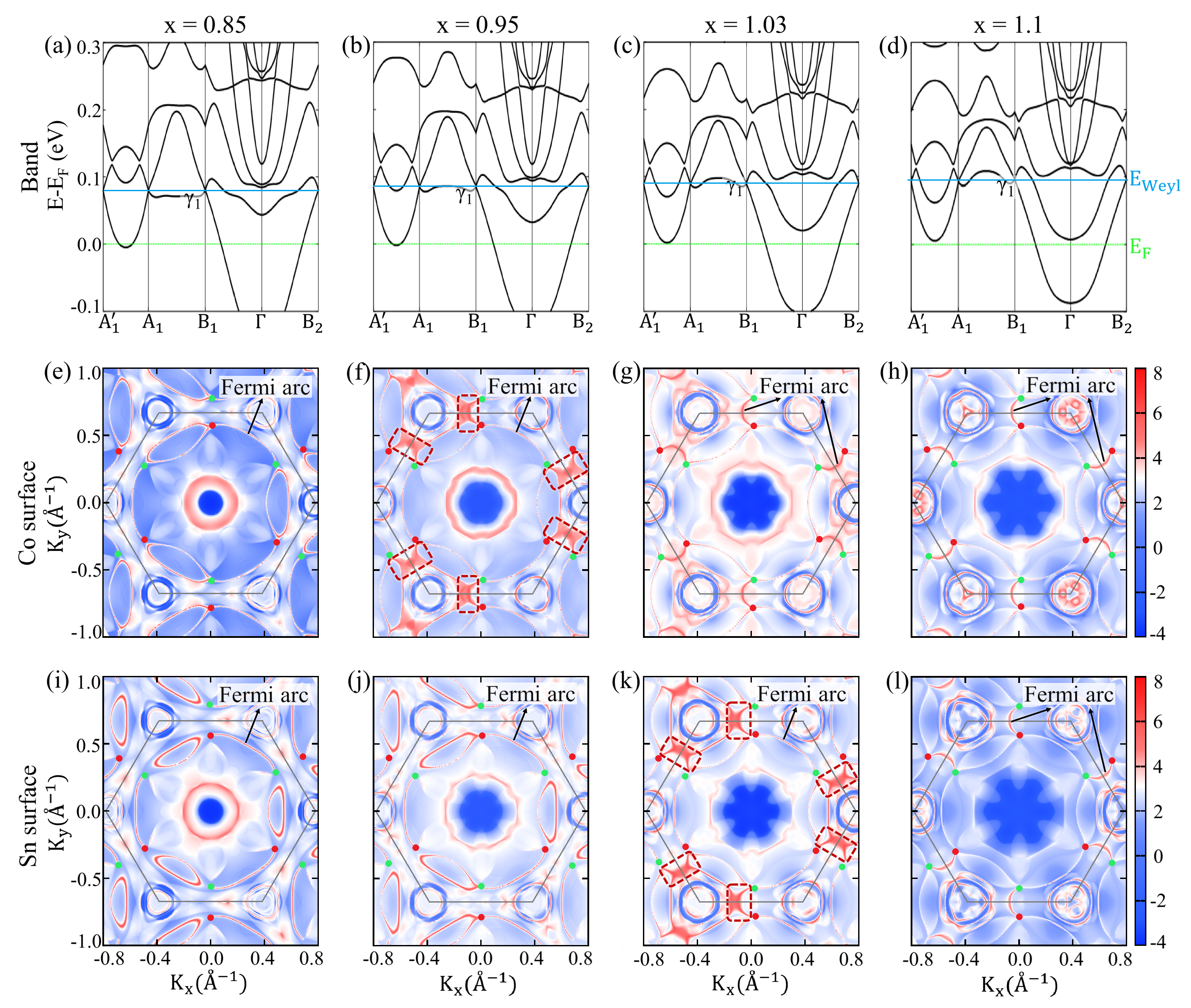}
	\caption{
		The band structure and Fermi arcs on the (001) surface for different terminations of Co$_3$Sn$_2$S$_2$ in the FM state.  The calculated band structure between different Weyl points (a-d), (001) surface states on Co (e-h) and Sn (i-l) terminated surfaces for $x$=0.85, 0.95, 1.03, and 1.1, respectively. $x$ denotes an effective electronic correlation strength. The pink and black points represent Weyl points with positive and negative chiralities, respectively. The dashed red box indicates the tendency of the Fermi arcs connecting the Weyl points between the adjacent Brillouin zones.}	
	\label{sf2}
\end{figure*}

The surface-dependent reconfiguration of the connectivity and topology of Fermi arcs in the DFT and DFT+DMFT calculations indicates that electronic correlation plays an importance role in modulating topological boundary states.
To explore how the Fermi arcs evolve with the strength of electronic correlation, we employ a linear combination of the DFT+DMFT Hamiltonian $H_{DFT+DMFT}$ and DFT Hamiltonian $H_{DFT}$ to build a model Hamiltonian $H(x)=x* H_{DFT+DMFT} + (1-x) * H_{DFT}$ where $x$ is used as a measurement of the effective electronic correlation strength ($x=0$ means weak electronic correlation as in DFT calculations while $x=1$ corresponds to moderate strength of electronic correlation as in the DFT+DMFT calculations). 
$H(x)$ is then used to compute the band structure and Fermi arcs on the Co and Sn-terminated surfaces, which are shown in Fig.~\ref{sf2} for $x$=0.85, 0.95, 1.03 and 1.1.

Progressive increase of the DFT+DMFT weighting ($x$) triggers critical phase transitions.
When $x$ = 0.95 (shown in Fig.~\ref{sf2}(b,f,j)), the transition states (the traces of Fermi arcs connects the Weyl points between adjacent BZs, as indicated by the dotted red box in Fig.~\ref{sf2}(f)) appears on Co surface, and the maximum energy of $\gamma_{1}$ band is equivalent to the energy of the Weyl points. while the Fermi arcs retain intra-BZ connectivity on Sn surface. 
As we further increase $x$ to 1.03 (shown in Fig.~\ref{sf2}(c,g,k)), the Fermi arcs on Co surface transform into connecting between adjacent BZs, and the transition states appear on Sn surface.  
With continuous increase in $\gamma_{1}$ band energy,when $x$ = 1.1 (shown in Fig.~\ref{sf2}(d,h,l)), the Fermi arcs on Co and Sn surfaces both transform into connecting the Weyl points between adjacent BZs.

According to the work by Chen Fang and collaborators\cite{nphys3782}, the nontrivial Chern number of Weyl points imposes a chiral signature on the surface dispersion along closed paths encircling their projections. 
This chirality drives the emergence of helicoid (anti-helicoid) structures near the Weyl points with positive (negative) chirality . 
The interplay between these opposing geometries generates open-shaped iso-energetic contours, known as Fermi arcs, which connect the projections of Weyl points on the surface BZ\cite{Wang04052017}.
Consequently, Fermi arcs are jointly governed by the spatial arrangement of Weyl points and the helicoidal geometry of surface states.
In our case, restructuring the $\gamma_{1}$ band modifies the energy-momentum relationship near Weyl points, altering the dispersion characteristics that governs the helicoidal geometry of surface states.
Concurrently, these band modifications shift the position of the Weyl points, thereby reconfiguring both the contour and connectivity of Fermi arcs.
Our calculated results are consistent with this mechanism. 
Prior to the appearance of Fermi arcs connecting Weyl points in adjacent Brillouin zones, analogous iso-energetic contours emerge in the helicoid projections (red dashed box in Fig.~\ref{sf2}(f) and (k)).
These transition signatures demonstrate that band dispersion relations predetermine the contour and connectivity of Fermi arcs.
\\

\section{Discussion}

In summary, our study highlights the significant role of electronic correlation in accounting for the electronic structure, charge dynamics and the topology and connectivity of Fermi arcs in Co$_3$Sn$_2$S$_2$. 
While the topological protection of connectivity between Weyl points of opposite chirality remains robust, their specific linkage patterns exhibit pronounced susceptibility to energies and locations of the Weyl points as well as the dispersion of the bands around the Weyl points. Tuning the electronic correlation strength can renormalize the bands and shift the Weyl points, thus manipulate the topology and connectivity of the Fermi arcs.

Fermi arcs have been shown to give rise to numerous intriguing phenomena, including unconventional Fermi-arc-mediated quantum oscillation, ultra-strong resonant peak, chiral magnetic effect and anisotropic optical conductivity\cite{yang2019topological,science.aad8766,PhysRevB.108.195436,PhysRevB.86.195102,jia2016weyl}. 
In this work, we demonstrate for the first time that electronic correlation fundamentally reshapes both the contour and connectivity of Fermi arcs in Weyl semimetals. 
The evolution of the Fermi arcs with varying electronic correlation strength reveals that band dispersion around the Weyl points predetermines the contour and connectivity of the Fermi arc, which establishes a predictive framework for designing topological surface states \textit{via} band-structure engineering. 
Our work will stimulate further investigation into manipulating the transport properties, novel metallic states, ultra quantum states, and Majorana zero modes \cite{PhysRevLett.130.196602,andp.201900449} of Weyl semimetals by tuning the Fermi arcs through band-structure engineering. 
\\

\section{Methods}

DFT calculations are carried out using the full-potential linear augmented plane wave method as implemented in WIEN2k\cite{10.1063/1.5143061}. 
The Perdew-Burke-Ernzerhof (PBE)\cite{PhysRevLett.77.3865} parametrization of the generalized gradient approximation is used for the exchange-correlation functional. 
We also carry out DFT calculations using the projector augmented wave pseudopotential method implemented in the Vienna ab initio simulation package (VASP)\cite{KRESSE199615, PhysRevB.54.11169}. 
The plane-wave cutoff energy is set to 500 eV. The results of the calculations from the two packages match very well.

To take in account electronic correlation effects, Co $3d$ electrons of the lattice problem are treated as impurities and mapped onto the Anderson impurity model. The quantum impurity problem is solved using the continuous time quantum Monte Carlo (CTQMC) method\cite{PhysRevB.75.155113, PhysRevLett.97.076405}. 
In the main text, $U$=4.0 eV and Hund's exchange $J_{H}$=0.8 eV are used in both the PM and FM states, while in the supplementary material, $U$=6.0 eV and Hund's exchange $J_{H}$=0.9 eV are employed. 
The DFT+DMFT calculations are carried out at 232 K in the PM state, and at 116 K in the FM state. 
The electronic charge is computed self-consistently on DFT+DMFT density matrix\cite{PhysRevB.81.195107}. 
The fully localized formula $U(n_{d}^{0}-1/2)-J(n_{d}^{0}-1)/2$ is used for the double-counting term, where $n_{d}^{0}$ is the nominal occupation of the Co $3d$ orbitals.

For calculating the surface states on the $(001)$ surface of Co$_3$Sn$_2$S$_2$, the DFT tight-binding Hamiltonian is obtained by maximally localized Wannier functions\cite{PhysRevB.56.12847, PhysRevB.65.035109, RevModPhys.84.1419} implemented in the WANNIER90 code\cite{MOSTOFI20142309}. 
Based on DFT tight-binding Hamiltonian, DFT+DMFT tight-binding Hamiltonian is obtained by modifying the hopping parameters to match the DFT+DMFT band structure around E$_F$\cite{PhysRevB.65.035109, MOSTOFI20142309}. 
Then based on DFT and DFT+DMFT tight-binding Hamiltonian, we compute the surface states by using WANNIERTOOLS package\cite{RevModPhys.84.1419, WU2018405}. 
The experimental crystal structure (space group $R\overline{3}m$, $No.$ 166) of Co$_3$Sn$_2$S$_2$ with hexagonal lattice constants $a$=$b$=5.3689 $\AA$ and $c$=13.176 $\AA$ are used in the calculations.
\\

\textbf{Data availability} \\
All raw data generated during the current study are available from the corresponding author upon request.

\textbf{Code availability} \\
The codes used for the DFT and DFT+DMFT calculations in this study are available from the corresponding authors upon request.

\begin{acknowledgments}
This work was supported by the Fundamental Research Funds for the Central Universities (Grant No. 2243300003),  Innovation Program for Quantum Science and Technology (2021ZD0302800), and the National Natural Science Foundation of China (Grant No. 12074041). Part of the calculations were carried out using the high performance computing cluster of Beijing Normal University in Zhuhai.
\end{acknowledgments}

\textbf{Author contributions} \\
Z.P.Y. conceived the research. Y. R. P. carried out the DFT and DFT+DMFT calculations. Y. R. P. and Z. P. Y. analyzed the results and wrote the paper. All authors participated in the discussion and comment on the paper. \\

\textbf{Competing interests} \\
The authors declare no competing interests. \\

\textbf{Additional information} \\
\textbf{Supplementary information} The online version contains supplementary material available at https://doi.org/(to be inserted) \\

\textbf{Correspondence} and requests for materials should be addressed to Zhiping Yin


\begin{thebibliography}{78}%
	\makeatletter
	\providecommand \@ifxundefined [1]{%
		\@ifx{#1\undefined}
	}%
	\providecommand \@ifnum [1]{%
		\ifnum #1\expandafter \@firstoftwo
		\else \expandafter \@secondoftwo
		\fi
	}%
	\providecommand \@ifx [1]{%
		\ifx #1\expandafter \@firstoftwo
		\else \expandafter \@secondoftwo
		\fi
	}%
	\providecommand \natexlab [1]{#1}%
	\providecommand \enquote  [1]{``#1''}%
	\providecommand \bibnamefont  [1]{#1}%
	\providecommand \bibfnamefont [1]{#1}%
	\providecommand \citenamefont [1]{#1}%
	\providecommand \href@noop [0]{\@secondoftwo}%
	\providecommand \href [0]{\begingroup \@sanitize@url \@href}%
	\providecommand \@href[1]{\@@startlink{#1}\@@href}%
	\providecommand \@@href[1]{\endgroup#1\@@endlink}%
	\providecommand \@sanitize@url [0]{\catcode `\\12\catcode `\$12\catcode
		`\&12\catcode `\#12\catcode `\^12\catcode `\_12\catcode `\%12\relax}%
	\providecommand \@@startlink[1]{}%
	\providecommand \@@endlink[0]{}%
	\providecommand \url  [0]{\begingroup\@sanitize@url \@url }%
	\providecommand \@url [1]{\endgroup\@href {#1}{\urlprefix }}%
	\providecommand \urlprefix  [0]{URL }%
	\providecommand \Eprint [0]{\href }%
	\providecommand \doibase [0]{https://doi.org/}%
	\providecommand \selectlanguage [0]{\@gobble}%
	\providecommand \bibinfo  [0]{\@secondoftwo}%
	\providecommand \bibfield  [0]{\@secondoftwo}%
	\providecommand \translation [1]{[#1]}%
	\providecommand \BibitemOpen [0]{}%
	\providecommand \bibitemStop [0]{}%
	\providecommand \bibitemNoStop [0]{.\EOS\space}%
	\providecommand \EOS [0]{\spacefactor3000\relax}%
	\providecommand \BibitemShut  [1]{\csname bibitem#1\endcsname}%
	\let\auto@bib@innerbib\@empty
	\bibitem [{\citenamefont {Chen}\ \emph {et~al.}(2016)\citenamefont {Chen},
		\citenamefont {Lv}, \citenamefont {Luo}, \citenamefont {Lu}, \citenamefont
		{Pei}, \citenamefont {Lin}, \citenamefont {Han}, \citenamefont {Zhu},
		\citenamefont {Song},\ and\ \citenamefont {Sun}}]{PhysRevB.94.235154}%
	\BibitemOpen
	\bibfield  {author} {\bibinfo {author} {\bibfnamefont {F.~C.}\ \bibnamefont
			{Chen}}, \bibinfo {author} {\bibfnamefont {H.~Y.}\ \bibnamefont {Lv}},
		\bibinfo {author} {\bibfnamefont {X.}~\bibnamefont {Luo}}, \bibinfo {author}
		{\bibfnamefont {W.~J.}\ \bibnamefont {Lu}}, \bibinfo {author} {\bibfnamefont
			{Q.~L.}\ \bibnamefont {Pei}}, \bibinfo {author} {\bibfnamefont {G.~T.}\
			\bibnamefont {Lin}}, \bibinfo {author} {\bibfnamefont {Y.~Y.}\ \bibnamefont
			{Han}}, \bibinfo {author} {\bibfnamefont {X.~B.}\ \bibnamefont {Zhu}},
		\bibinfo {author} {\bibfnamefont {W.~H.}\ \bibnamefont {Song}},\ and\
		\bibinfo {author} {\bibfnamefont {Y.~P.}\ \bibnamefont {Sun}},\ }\bibfield
	{title} {\bibinfo {title} {Extremely large magnetoresistance in the
			type-\uppercase\expandafter{\romannumeral2} weyl semimetal
			$\mathrm{MoTe}_{2}$},\ }\href {https://doi.org/10.1103/PhysRevB.94.235154}
	{\bibfield  {journal} {\bibinfo  {journal} {Phys. Rev. B}\ }\textbf {\bibinfo
			{volume} {94}},\ \bibinfo {pages} {235154} (\bibinfo {year}
		{2016})}\BibitemShut {NoStop}%
	\bibitem [{\citenamefont {Liu}\ \emph {et~al.}(2018)\citenamefont {Liu},
		\citenamefont {Sun}, \citenamefont {Kumar}, \citenamefont {Muechler},
		\citenamefont {Sun}, \citenamefont {Jiao}, \citenamefont {Yang},
		\citenamefont {Liu}, \citenamefont {Liang}, \citenamefont {Xu} \emph
		{et~al.}}]{liu2018giant}%
	\BibitemOpen
	\bibfield  {author} {\bibinfo {author} {\bibfnamefont {E.}~\bibnamefont
			{Liu}}, \bibinfo {author} {\bibfnamefont {Y.}~\bibnamefont {Sun}}, \bibinfo
		{author} {\bibfnamefont {N.}~\bibnamefont {Kumar}}, \bibinfo {author}
		{\bibfnamefont {L.}~\bibnamefont {Muechler}}, \bibinfo {author}
		{\bibfnamefont {A.}~\bibnamefont {Sun}}, \bibinfo {author} {\bibfnamefont
			{L.}~\bibnamefont {Jiao}}, \bibinfo {author} {\bibfnamefont {S.-Y.}\
			\bibnamefont {Yang}}, \bibinfo {author} {\bibfnamefont {D.}~\bibnamefont
			{Liu}}, \bibinfo {author} {\bibfnamefont {A.}~\bibnamefont {Liang}}, \bibinfo
		{author} {\bibfnamefont {Q.}~\bibnamefont {Xu}}, \emph {et~al.},\ }\bibfield
	{title} {\bibinfo {title} {Giant anomalous $\mathrm{Hall}$ effect in a
			ferromagnetic $\mathrm{Kagome}$-lattice semimetal},\ }\href
	{https://doi.org/10.1038/s41567-018-0234-5} {\bibfield  {journal} {\bibinfo
			{journal} {Nature Physics}\ }\textbf {\bibinfo {volume} {14}},\ \bibinfo
		{pages} {1125} (\bibinfo {year} {2018})}\BibitemShut {NoStop}%
	\bibitem [{\citenamefont {Nagpal}\ and\ \citenamefont
		{Patnaik}(2020)}]{Nagpal_2020}%
	\BibitemOpen
	\bibfield  {author} {\bibinfo {author} {\bibfnamefont {V.}~\bibnamefont
			{Nagpal}}\ and\ \bibinfo {author} {\bibfnamefont {S.}~\bibnamefont
			{Patnaik}},\ }\bibfield  {title} {\bibinfo {title} {Breakdown of
			$\mathrm{Ohm’s}$ law and nontrivial $\mathrm{Berry}$ phase in magnetic
			$\mathrm{Weyl}$ semimetal $\mathrm{Co}_{3}\mathrm{Sn}_{2}\mathrm{S}_{2}$},\
	}\href {https://doi.org/10.1088/1361-648X/ab9859} {\bibfield  {journal}
		{\bibinfo  {journal} {Journal of Physics: Condensed Matter}\ }\textbf
		{\bibinfo {volume} {32}},\ \bibinfo {pages} {405602} (\bibinfo {year}
		{2020})}\BibitemShut {NoStop}%
	\bibitem [{\citenamefont {Huang}\ \emph
		{et~al.}(2015{\natexlab{a}})\citenamefont {Huang}, \citenamefont {Zhao},
		\citenamefont {Long}, \citenamefont {Wang}, \citenamefont {Chen},
		\citenamefont {Yang}, \citenamefont {Liang}, \citenamefont {Xue},
		\citenamefont {Weng}, \citenamefont {Fang}, \citenamefont {Dai},\ and\
		\citenamefont {Chen}}]{PhysRevX.5.031023}%
	\BibitemOpen
	\bibfield  {author} {\bibinfo {author} {\bibfnamefont {X.}~\bibnamefont
			{Huang}}, \bibinfo {author} {\bibfnamefont {L.}~\bibnamefont {Zhao}},
		\bibinfo {author} {\bibfnamefont {Y.}~\bibnamefont {Long}}, \bibinfo {author}
		{\bibfnamefont {P.}~\bibnamefont {Wang}}, \bibinfo {author} {\bibfnamefont
			{D.}~\bibnamefont {Chen}}, \bibinfo {author} {\bibfnamefont {Z.}~\bibnamefont
			{Yang}}, \bibinfo {author} {\bibfnamefont {H.}~\bibnamefont {Liang}},
		\bibinfo {author} {\bibfnamefont {M.}~\bibnamefont {Xue}}, \bibinfo {author}
		{\bibfnamefont {H.}~\bibnamefont {Weng}}, \bibinfo {author} {\bibfnamefont
			{Z.}~\bibnamefont {Fang}}, \bibinfo {author} {\bibfnamefont {X.}~\bibnamefont
			{Dai}},\ and\ \bibinfo {author} {\bibfnamefont {G.}~\bibnamefont {Chen}},\
	}\bibfield  {title} {\bibinfo {title} {Observation of the
			$\mathrm{Chiral}$-$\mathrm{Anomaly}$-$\mathrm{Induced}$ $\mathrm{Negative}$
			$\mathrm{Magnetoresistance}$ in $\mathrm{3D}$ $\mathrm{Weyl}$
			$\mathrm{Semimetal}$ $\mathrm{TaAs}$},\ }\href
	{https://doi.org/10.1103/PhysRevX.5.031023} {\bibfield  {journal} {\bibinfo
			{journal} {Phys. Rev. X}\ }\textbf {\bibinfo {volume} {5}},\ \bibinfo {pages}
		{031023} (\bibinfo {year} {2015}{\natexlab{a}})}\BibitemShut {NoStop}%
	\bibitem [{\citenamefont {Zhang}\ \emph {et~al.}(2016)\citenamefont {Zhang},
		\citenamefont {Xu}, \citenamefont {Belopolski}, \citenamefont {Yuan},
		\citenamefont {Lin}, \citenamefont {Tong}, \citenamefont {Bian},
		\citenamefont {Alidoust}, \citenamefont {Lee}, \citenamefont {Huang} \emph
		{et~al.}}]{zhang2016signatures}%
	\BibitemOpen
	\bibfield  {author} {\bibinfo {author} {\bibfnamefont {C.-L.}\ \bibnamefont
			{Zhang}}, \bibinfo {author} {\bibfnamefont {S.-Y.}\ \bibnamefont {Xu}},
		\bibinfo {author} {\bibfnamefont {I.}~\bibnamefont {Belopolski}}, \bibinfo
		{author} {\bibfnamefont {Z.}~\bibnamefont {Yuan}}, \bibinfo {author}
		{\bibfnamefont {Z.}~\bibnamefont {Lin}}, \bibinfo {author} {\bibfnamefont
			{B.}~\bibnamefont {Tong}}, \bibinfo {author} {\bibfnamefont {G.}~\bibnamefont
			{Bian}}, \bibinfo {author} {\bibfnamefont {N.}~\bibnamefont {Alidoust}},
		\bibinfo {author} {\bibfnamefont {C.-C.}\ \bibnamefont {Lee}}, \bibinfo
		{author} {\bibfnamefont {S.-M.}\ \bibnamefont {Huang}}, \emph {et~al.},\
	}\bibfield  {title} {\bibinfo {title} {Signatures of the
			$\mathrm{Adler}$-$\mathrm{Bell}$-$\mathrm{Jackiw}$ chiral anomaly in a
			$\mathrm{Weyl}$ fermion semimetal},\ }\href
	{https://doi.org/10.1038/ncomms10735} {\bibfield  {journal} {\bibinfo
			{journal} {Nature Communications}\ }\textbf {\bibinfo {volume} {7}},\
		\bibinfo {pages} {1} (\bibinfo {year} {2016})}\BibitemShut {NoStop}%
	\bibitem [{\citenamefont {Wang}\ \emph
		{et~al.}(2016{\natexlab{a}})\citenamefont {Wang}, \citenamefont {Zheng},
		\citenamefont {Shen}, \citenamefont {Lu}, \citenamefont {Fang}, \citenamefont
		{Sheng}, \citenamefont {Zhou}, \citenamefont {Yang}, \citenamefont {Li},
		\citenamefont {Feng},\ and\ \citenamefont {Xu}}]{PhysRevB.93.121112}%
	\BibitemOpen
	\bibfield  {author} {\bibinfo {author} {\bibfnamefont {Z.}~\bibnamefont
			{Wang}}, \bibinfo {author} {\bibfnamefont {Y.}~\bibnamefont {Zheng}},
		\bibinfo {author} {\bibfnamefont {Z.}~\bibnamefont {Shen}}, \bibinfo {author}
		{\bibfnamefont {Y.}~\bibnamefont {Lu}}, \bibinfo {author} {\bibfnamefont
			{H.}~\bibnamefont {Fang}}, \bibinfo {author} {\bibfnamefont {F.}~\bibnamefont
			{Sheng}}, \bibinfo {author} {\bibfnamefont {Y.}~\bibnamefont {Zhou}},
		\bibinfo {author} {\bibfnamefont {X.}~\bibnamefont {Yang}}, \bibinfo {author}
		{\bibfnamefont {Y.}~\bibnamefont {Li}}, \bibinfo {author} {\bibfnamefont
			{C.}~\bibnamefont {Feng}},\ and\ \bibinfo {author} {\bibfnamefont {Z.-A.}\
			\bibnamefont {Xu}},\ }\bibfield  {title} {\bibinfo {title}
		{Helicity-protected ultrahigh mobility $\mathrm{Weyl}$ fermions in
			$\mathrm{NbP}$},\ }\href {https://doi.org/10.1103/PhysRevB.93.121112}
	{\bibfield  {journal} {\bibinfo  {journal} {Phys. Rev. B}\ }\textbf {\bibinfo
			{volume} {93}},\ \bibinfo {pages} {121112} (\bibinfo {year}
		{2016}{\natexlab{a}})}\BibitemShut {NoStop}%
	\bibitem [{\citenamefont {Niemann}\ \emph {et~al.}(2017)\citenamefont
		{Niemann}, \citenamefont {Gooth}, \citenamefont {Wu}, \citenamefont
		{B{\"a}{\ss}ler}, \citenamefont {Sergelius}, \citenamefont {H{\"u}hne},
		\citenamefont {Rellinghaus}, \citenamefont {Shekhar}, \citenamefont
		{S{\"u}{\ss}}, \citenamefont {Schmidt} \emph {et~al.}}]{niemann2017chiral}%
	\BibitemOpen
	\bibfield  {author} {\bibinfo {author} {\bibfnamefont {A.~C.}\ \bibnamefont
			{Niemann}}, \bibinfo {author} {\bibfnamefont {J.}~\bibnamefont {Gooth}},
		\bibinfo {author} {\bibfnamefont {S.-C.}\ \bibnamefont {Wu}}, \bibinfo
		{author} {\bibfnamefont {S.}~\bibnamefont {B{\"a}{\ss}ler}}, \bibinfo
		{author} {\bibfnamefont {P.}~\bibnamefont {Sergelius}}, \bibinfo {author}
		{\bibfnamefont {R.}~\bibnamefont {H{\"u}hne}}, \bibinfo {author}
		{\bibfnamefont {B.}~\bibnamefont {Rellinghaus}}, \bibinfo {author}
		{\bibfnamefont {C.}~\bibnamefont {Shekhar}}, \bibinfo {author} {\bibfnamefont
			{V.}~\bibnamefont {S{\"u}{\ss}}}, \bibinfo {author} {\bibfnamefont
			{M.}~\bibnamefont {Schmidt}}, \emph {et~al.},\ }\bibfield  {title} {\bibinfo
		{title} {Chiral magnetoresistance in the $\mathrm{Weyl}$ semimetal
			$\mathrm{NbP}$},\ }\href {https://doi.org/10.1038/srep43394} {\bibfield
		{journal} {\bibinfo  {journal} {Scientific Reports}\ }\textbf {\bibinfo
			{volume} {7}},\ \bibinfo {pages} {43394} (\bibinfo {year}
		{2017})}\BibitemShut {NoStop}%
	\bibitem [{\citenamefont {Gooth}\ \emph {et~al.}(2017)\citenamefont {Gooth},
		\citenamefont {Niemann}, \citenamefont {Meng}, \citenamefont {Grushin},
		\citenamefont {Landsteiner}, \citenamefont {Gotsmann}, \citenamefont
		{Menges}, \citenamefont {Schmidt}, \citenamefont {Shekhar}, \citenamefont
		{S{\"u}{\ss}} \emph {et~al.}}]{gooth2017experimental}%
	\BibitemOpen
	\bibfield  {author} {\bibinfo {author} {\bibfnamefont {J.}~\bibnamefont
			{Gooth}}, \bibinfo {author} {\bibfnamefont {A.~C.}\ \bibnamefont {Niemann}},
		\bibinfo {author} {\bibfnamefont {T.}~\bibnamefont {Meng}}, \bibinfo {author}
		{\bibfnamefont {A.~G.}\ \bibnamefont {Grushin}}, \bibinfo {author}
		{\bibfnamefont {K.}~\bibnamefont {Landsteiner}}, \bibinfo {author}
		{\bibfnamefont {B.}~\bibnamefont {Gotsmann}}, \bibinfo {author}
		{\bibfnamefont {F.}~\bibnamefont {Menges}}, \bibinfo {author} {\bibfnamefont
			{M.}~\bibnamefont {Schmidt}}, \bibinfo {author} {\bibfnamefont
			{C.}~\bibnamefont {Shekhar}}, \bibinfo {author} {\bibfnamefont
			{V.}~\bibnamefont {S{\"u}{\ss}}}, \emph {et~al.},\ }\bibfield  {title}
	{\bibinfo {title} {Experimental signatures of the mixed axial-gravitational
			anomaly in the $\mathrm{Weyl}$ semimetal $\mathrm{NbP}$},\ }\href
	{https://doi.org/10.1038/nature23005} {\bibfield  {journal} {\bibinfo
			{journal} {Nature}\ }\textbf {\bibinfo {volume} {547}},\ \bibinfo {pages}
		{324} (\bibinfo {year} {2017})}\BibitemShut {NoStop}%
	\bibitem [{\citenamefont {Sun}\ \emph {et~al.}(2016)\citenamefont {Sun},
		\citenamefont {Zhang}, \citenamefont {Felser},\ and\ \citenamefont
		{Yan}}]{PhysRevLett.117.146403}%
	\BibitemOpen
	\bibfield  {author} {\bibinfo {author} {\bibfnamefont {Y.}~\bibnamefont
			{Sun}}, \bibinfo {author} {\bibfnamefont {Y.}~\bibnamefont {Zhang}}, \bibinfo
		{author} {\bibfnamefont {C.}~\bibnamefont {Felser}},\ and\ \bibinfo {author}
		{\bibfnamefont {B.}~\bibnamefont {Yan}},\ }\bibfield  {title} {\bibinfo
		{title} {Strong $\mathrm{Intrinsic}$ $\mathrm{Spin}$ $\mathrm{Hall}$
			$\mathrm{Effect}$ in the $\mathrm{TaAs}$ $\mathrm{Family}$ of $\mathrm{Weyl}$
			$\mathrm{Semimetals}$},\ }\href
	{https://doi.org/10.1103/PhysRevLett.117.146403} {\bibfield  {journal}
		{\bibinfo  {journal} {Phys. Rev. Lett.}\ }\textbf {\bibinfo {volume} {117}},\
		\bibinfo {pages} {146403} (\bibinfo {year} {2016})}\BibitemShut {NoStop}%
	\bibitem [{\citenamefont {Xu}\ \emph {et~al.}(2011)\citenamefont {Xu},
		\citenamefont {Weng}, \citenamefont {Wang}, \citenamefont {Dai},\ and\
		\citenamefont {Fang}}]{PhysRevLett.107.186806}%
	\BibitemOpen
	\bibfield  {author} {\bibinfo {author} {\bibfnamefont {G.}~\bibnamefont
			{Xu}}, \bibinfo {author} {\bibfnamefont {H.}~\bibnamefont {Weng}}, \bibinfo
		{author} {\bibfnamefont {Z.}~\bibnamefont {Wang}}, \bibinfo {author}
		{\bibfnamefont {X.}~\bibnamefont {Dai}},\ and\ \bibinfo {author}
		{\bibfnamefont {Z.}~\bibnamefont {Fang}},\ }\bibfield  {title} {\bibinfo
		{title} {Chern $\mathrm{Semimetal}$ and the $\mathrm{Quantized}$
			$\mathrm{Anomalous}$ $\mathrm{Hall}$ $\mathrm{Effect}$ in
			$\mathrm{HgCr}_{2}\mathrm{Se}_{4}$},\ }\href
	{https://doi.org/10.1103/PhysRevLett.107.186806} {\bibfield  {journal}
		{\bibinfo  {journal} {Phys. Rev. Lett.}\ }\textbf {\bibinfo {volume} {107}},\
		\bibinfo {pages} {186806} (\bibinfo {year} {2011})}\BibitemShut {NoStop}%
	\bibitem [{\citenamefont {Burkov}\ and\ \citenamefont
		{Balents}(2011)}]{PhysRevLett.107.127205}%
	\BibitemOpen
	\bibfield  {author} {\bibinfo {author} {\bibfnamefont {A.~A.}\ \bibnamefont
			{Burkov}}\ and\ \bibinfo {author} {\bibfnamefont {L.}~\bibnamefont
			{Balents}},\ }\bibfield  {title} {\bibinfo {title} {Weyl $\mathrm{Semimetal}$
			in a $\mathrm{Topological}$ $\mathrm{Insulator}$ $\mathrm{Multilayer}$},\
	}\href {https://doi.org/10.1103/PhysRevLett.107.127205} {\bibfield  {journal}
		{\bibinfo  {journal} {Phys. Rev. Lett.}\ }\textbf {\bibinfo {volume} {107}},\
		\bibinfo {pages} {127205} (\bibinfo {year} {2011})}\BibitemShut {NoStop}%
	\bibitem [{\citenamefont {Weng}\ \emph {et~al.}(2015)\citenamefont {Weng},
		\citenamefont {Fang}, \citenamefont {Fang}, \citenamefont {Bernevig},\ and\
		\citenamefont {Dai}}]{PhysRevX.5.011029}%
	\BibitemOpen
	\bibfield  {author} {\bibinfo {author} {\bibfnamefont {H.}~\bibnamefont
			{Weng}}, \bibinfo {author} {\bibfnamefont {C.}~\bibnamefont {Fang}}, \bibinfo
		{author} {\bibfnamefont {Z.}~\bibnamefont {Fang}}, \bibinfo {author}
		{\bibfnamefont {B.~A.}\ \bibnamefont {Bernevig}},\ and\ \bibinfo {author}
		{\bibfnamefont {X.}~\bibnamefont {Dai}},\ }\bibfield  {title} {\bibinfo
		{title} {Weyl $\mathrm{Semimetal}$ $\mathrm{Phase}$ in
			$\mathrm{Noncentrosymmetric}$ $\mathrm{Transition}$-$\mathrm{Metal}$
			$\mathrm{Monophosphides}$},\ }\href
	{https://doi.org/10.1103/PhysRevX.5.011029} {\bibfield  {journal} {\bibinfo
			{journal} {Phys. Rev. X}\ }\textbf {\bibinfo {volume} {5}},\ \bibinfo {pages}
		{011029} (\bibinfo {year} {2015})}\BibitemShut {NoStop}%
	\bibitem [{\citenamefont {Huang}\ \emph
		{et~al.}(2015{\natexlab{b}})\citenamefont {Huang}, \citenamefont {Xu},
		\citenamefont {Belopolski}, \citenamefont {Lee}, \citenamefont {Chang},
		\citenamefont {Wang}, \citenamefont {Alidoust}, \citenamefont {Bian},
		\citenamefont {Neupane}, \citenamefont {Zhang} \emph
		{et~al.}}]{huang2015weyl}%
	\BibitemOpen
	\bibfield  {author} {\bibinfo {author} {\bibfnamefont {S.-M.}\ \bibnamefont
			{Huang}}, \bibinfo {author} {\bibfnamefont {S.-Y.}\ \bibnamefont {Xu}},
		\bibinfo {author} {\bibfnamefont {I.}~\bibnamefont {Belopolski}}, \bibinfo
		{author} {\bibfnamefont {C.-C.}\ \bibnamefont {Lee}}, \bibinfo {author}
		{\bibfnamefont {G.}~\bibnamefont {Chang}}, \bibinfo {author} {\bibfnamefont
			{B.}~\bibnamefont {Wang}}, \bibinfo {author} {\bibfnamefont {N.}~\bibnamefont
			{Alidoust}}, \bibinfo {author} {\bibfnamefont {G.}~\bibnamefont {Bian}},
		\bibinfo {author} {\bibfnamefont {M.}~\bibnamefont {Neupane}}, \bibinfo
		{author} {\bibfnamefont {C.}~\bibnamefont {Zhang}}, \emph {et~al.},\
	}\bibfield  {title} {\bibinfo {title} {A $\mathrm{Weyl}$ $\mathrm{Fermion}$
			semimetal with surface $\mathrm{Fermi}$ arcs in the transition metal
			monopnictide $\mathrm{TaAs}$ class},\ }\href
	{https://doi.org/10.1038/ncomms8373} {\bibfield  {journal} {\bibinfo
			{journal} {Nature Communications}\ }\textbf {\bibinfo {volume} {6}},\
		\bibinfo {pages} {7373} (\bibinfo {year} {2015}{\natexlab{b}})}\BibitemShut
	{NoStop}%
	\bibitem [{\citenamefont {Xu}\ \emph {et~al.}(2015)\citenamefont {Xu},
		\citenamefont {Belopolski}, \citenamefont {Alidoust}, \citenamefont
		{Neupane}, \citenamefont {Bian}, \citenamefont {Zhang}, \citenamefont
		{Sankar}, \citenamefont {Chang}, \citenamefont {Yuan}, \citenamefont {Lee},
		\citenamefont {Huang}, \citenamefont {Zheng}, \citenamefont {Ma},
		\citenamefont {Sanchez}, \citenamefont {Wang}, \citenamefont {Bansil},
		\citenamefont {Chou}, \citenamefont {Shibayev}, \citenamefont {Lin},
		\citenamefont {Jia},\ and\ \citenamefont {Hasan}}]{science.aaa9297}%
	\BibitemOpen
	\bibfield  {author} {\bibinfo {author} {\bibfnamefont {S.-Y.}\ \bibnamefont
			{Xu}}, \bibinfo {author} {\bibfnamefont {I.}~\bibnamefont {Belopolski}},
		\bibinfo {author} {\bibfnamefont {N.}~\bibnamefont {Alidoust}}, \bibinfo
		{author} {\bibfnamefont {M.}~\bibnamefont {Neupane}}, \bibinfo {author}
		{\bibfnamefont {G.}~\bibnamefont {Bian}}, \bibinfo {author} {\bibfnamefont
			{C.}~\bibnamefont {Zhang}}, \bibinfo {author} {\bibfnamefont
			{R.}~\bibnamefont {Sankar}}, \bibinfo {author} {\bibfnamefont
			{G.}~\bibnamefont {Chang}}, \bibinfo {author} {\bibfnamefont
			{Z.}~\bibnamefont {Yuan}}, \bibinfo {author} {\bibfnamefont {C.-C.}\
			\bibnamefont {Lee}}, \bibinfo {author} {\bibfnamefont {S.-M.}\ \bibnamefont
			{Huang}}, \bibinfo {author} {\bibfnamefont {H.}~\bibnamefont {Zheng}},
		\bibinfo {author} {\bibfnamefont {J.}~\bibnamefont {Ma}}, \bibinfo {author}
		{\bibfnamefont {D.~S.}\ \bibnamefont {Sanchez}}, \bibinfo {author}
		{\bibfnamefont {B.}~\bibnamefont {Wang}}, \bibinfo {author} {\bibfnamefont
			{A.}~\bibnamefont {Bansil}}, \bibinfo {author} {\bibfnamefont
			{F.}~\bibnamefont {Chou}}, \bibinfo {author} {\bibfnamefont {P.~P.}\
			\bibnamefont {Shibayev}}, \bibinfo {author} {\bibfnamefont {H.}~\bibnamefont
			{Lin}}, \bibinfo {author} {\bibfnamefont {S.}~\bibnamefont {Jia}},\ and\
		\bibinfo {author} {\bibfnamefont {M.~Z.}\ \bibnamefont {Hasan}},\ }\bibfield
	{title} {\bibinfo {title} {Discovery of a $\mathrm{Weyl}$ fermion semimetal
			and topological $\mathrm{Fermi}$ arcs},\ }\href
	{https://doi.org/10.1126/science.aaa9297} {\bibfield  {journal} {\bibinfo
			{journal} {Science}\ }\textbf {\bibinfo {volume} {349}},\ \bibinfo {pages}
		{613} (\bibinfo {year} {2015})}\BibitemShut {NoStop}%
	\bibitem [{\citenamefont {Lv}\ \emph {et~al.}(2015)\citenamefont {Lv},
		\citenamefont {Weng}, \citenamefont {Fu}, \citenamefont {Wang}, \citenamefont
		{Miao}, \citenamefont {Ma}, \citenamefont {Richard}, \citenamefont {Huang},
		\citenamefont {Zhao}, \citenamefont {Chen}, \citenamefont {Fang},
		\citenamefont {Dai}, \citenamefont {Qian},\ and\ \citenamefont
		{Ding}}]{PhysRevX.5.031013}%
	\BibitemOpen
	\bibfield  {author} {\bibinfo {author} {\bibfnamefont {B.~Q.}\ \bibnamefont
			{Lv}}, \bibinfo {author} {\bibfnamefont {H.~M.}\ \bibnamefont {Weng}},
		\bibinfo {author} {\bibfnamefont {B.~B.}\ \bibnamefont {Fu}}, \bibinfo
		{author} {\bibfnamefont {X.~P.}\ \bibnamefont {Wang}}, \bibinfo {author}
		{\bibfnamefont {H.}~\bibnamefont {Miao}}, \bibinfo {author} {\bibfnamefont
			{J.}~\bibnamefont {Ma}}, \bibinfo {author} {\bibfnamefont {P.}~\bibnamefont
			{Richard}}, \bibinfo {author} {\bibfnamefont {X.~C.}\ \bibnamefont {Huang}},
		\bibinfo {author} {\bibfnamefont {L.~X.}\ \bibnamefont {Zhao}}, \bibinfo
		{author} {\bibfnamefont {G.~F.}\ \bibnamefont {Chen}}, \bibinfo {author}
		{\bibfnamefont {Z.}~\bibnamefont {Fang}}, \bibinfo {author} {\bibfnamefont
			{X.}~\bibnamefont {Dai}}, \bibinfo {author} {\bibfnamefont {T.}~\bibnamefont
			{Qian}},\ and\ \bibinfo {author} {\bibfnamefont {H.}~\bibnamefont {Ding}},\
	}\bibfield  {title} {\bibinfo {title} {Experimental $\mathrm{Discovery}$ of
			$\mathrm{Weyl}$ $\mathrm{Semimetal}$ $\mathrm{TaAs}$},\ }\href
	{https://doi.org/10.1103/PhysRevX.5.031013} {\bibfield  {journal} {\bibinfo
			{journal} {Phys. Rev. X}\ }\textbf {\bibinfo {volume} {5}},\ \bibinfo {pages}
		{031013} (\bibinfo {year} {2015})}\BibitemShut {NoStop}%
	\bibitem [{\citenamefont {Yang}\ \emph {et~al.}(2015)\citenamefont {Yang},
		\citenamefont {Liu}, \citenamefont {Sun}, \citenamefont {Peng}, \citenamefont
		{Yang}, \citenamefont {Zhang}, \citenamefont {Zhou}, \citenamefont {Zhang},
		\citenamefont {Guo}, \citenamefont {Rahn} \emph {et~al.}}]{yang2015weyl}%
	\BibitemOpen
	\bibfield  {author} {\bibinfo {author} {\bibfnamefont {L.}~\bibnamefont
			{Yang}}, \bibinfo {author} {\bibfnamefont {Z.}~\bibnamefont {Liu}}, \bibinfo
		{author} {\bibfnamefont {Y.}~\bibnamefont {Sun}}, \bibinfo {author}
		{\bibfnamefont {H.}~\bibnamefont {Peng}}, \bibinfo {author} {\bibfnamefont
			{H.}~\bibnamefont {Yang}}, \bibinfo {author} {\bibfnamefont {T.}~\bibnamefont
			{Zhang}}, \bibinfo {author} {\bibfnamefont {B.}~\bibnamefont {Zhou}},
		\bibinfo {author} {\bibfnamefont {Y.}~\bibnamefont {Zhang}}, \bibinfo
		{author} {\bibfnamefont {Y.}~\bibnamefont {Guo}}, \bibinfo {author}
		{\bibfnamefont {M.}~\bibnamefont {Rahn}}, \emph {et~al.},\ }\bibfield
	{title} {\bibinfo {title} {Weyl semimetal phase in the non-centrosymmetric
			compound $\mathrm{TaAs}$},\ }\href {https://doi.org/10.1038/nphys3425}
	{\bibfield  {journal} {\bibinfo  {journal} {Nature Physics}\ }\textbf
		{\bibinfo {volume} {11}},\ \bibinfo {pages} {728} (\bibinfo {year}
		{2015})}\BibitemShut {NoStop}%
	\bibitem [{\citenamefont {Wan}\ \emph {et~al.}(2011)\citenamefont {Wan},
		\citenamefont {Turner}, \citenamefont {Vishwanath},\ and\ \citenamefont
		{Savrasov}}]{PhysRevB.83.205101}%
	\BibitemOpen
	\bibfield  {author} {\bibinfo {author} {\bibfnamefont {X.}~\bibnamefont
			{Wan}}, \bibinfo {author} {\bibfnamefont {A.~M.}\ \bibnamefont {Turner}},
		\bibinfo {author} {\bibfnamefont {A.}~\bibnamefont {Vishwanath}},\ and\
		\bibinfo {author} {\bibfnamefont {S.~Y.}\ \bibnamefont {Savrasov}},\
	}\bibfield  {title} {\bibinfo {title} {Topological semimetal and
			$\mathrm{Fermi}$-arc surface states in the electronic structure of pyrochlore
			iridates},\ }\href {https://doi.org/10.1103/PhysRevB.83.205101} {\bibfield
		{journal} {\bibinfo  {journal} {Phys. Rev. B}\ }\textbf {\bibinfo {volume}
			{83}},\ \bibinfo {pages} {205101} (\bibinfo {year} {2011})}\BibitemShut
	{NoStop}%
	\bibitem [{\citenamefont {Armitage}\ \emph {et~al.}(2018)\citenamefont
		{Armitage}, \citenamefont {Mele},\ and\ \citenamefont
		{Vishwanath}}]{RevModPhys.90.015001}%
	\BibitemOpen
	\bibfield  {author} {\bibinfo {author} {\bibfnamefont {N.~P.}\ \bibnamefont
			{Armitage}}, \bibinfo {author} {\bibfnamefont {E.~J.}\ \bibnamefont {Mele}},\
		and\ \bibinfo {author} {\bibfnamefont {A.}~\bibnamefont {Vishwanath}},\
	}\bibfield  {title} {\bibinfo {title} {Weyl and $\mathrm{Dirac}$ semimetals
			in three-dimensional solids},\ }\href
	{https://doi.org/10.1103/RevModPhys.90.015001} {\bibfield  {journal}
		{\bibinfo  {journal} {Rev. Mod. Phys.}\ }\textbf {\bibinfo {volume} {90}},\
		\bibinfo {pages} {015001} (\bibinfo {year} {2018})}\BibitemShut {NoStop}%
	\bibitem [{\citenamefont {Wang}\ \emph
		{et~al.}(2016{\natexlab{b}})\citenamefont {Wang}, \citenamefont {Vergniory},
		\citenamefont {Kushwaha}, \citenamefont {Hirschberger}, \citenamefont
		{Chulkov}, \citenamefont {Ernst}, \citenamefont {Ong}, \citenamefont {Cava},\
		and\ \citenamefont {Bernevig}}]{PhysRevLett.117.236401}%
	\BibitemOpen
	\bibfield  {author} {\bibinfo {author} {\bibfnamefont {Z.}~\bibnamefont
			{Wang}}, \bibinfo {author} {\bibfnamefont {M.~G.}\ \bibnamefont {Vergniory}},
		\bibinfo {author} {\bibfnamefont {S.}~\bibnamefont {Kushwaha}}, \bibinfo
		{author} {\bibfnamefont {M.}~\bibnamefont {Hirschberger}}, \bibinfo {author}
		{\bibfnamefont {E.~V.}\ \bibnamefont {Chulkov}}, \bibinfo {author}
		{\bibfnamefont {A.}~\bibnamefont {Ernst}}, \bibinfo {author} {\bibfnamefont
			{N.~P.}\ \bibnamefont {Ong}}, \bibinfo {author} {\bibfnamefont {R.~J.}\
			\bibnamefont {Cava}},\ and\ \bibinfo {author} {\bibfnamefont {B.~A.}\
			\bibnamefont {Bernevig}},\ }\bibfield  {title} {\bibinfo {title}
		{$\mathrm{Time}$-$\mathrm{Reversal}$-$\mathrm{Breaking}$ $\mathrm{Weyl}$
			$\mathrm{Fermions}$ in $\mathrm{Magnetic}$ $\mathrm{Heusler}$
			$\mathrm{Alloys}$},\ }\href {https://doi.org/10.1103/PhysRevLett.117.236401}
	{\bibfield  {journal} {\bibinfo  {journal} {Phys. Rev. Lett.}\ }\textbf
		{\bibinfo {volume} {117}},\ \bibinfo {pages} {236401} (\bibinfo {year}
		{2016}{\natexlab{b}})}\BibitemShut {NoStop}%
	\bibitem [{\citenamefont {Sun}\ \emph {et~al.}(2015)\citenamefont {Sun},
		\citenamefont {Wu},\ and\ \citenamefont {Yan}}]{PhysRevB.92.115428}%
	\BibitemOpen
	\bibfield  {author} {\bibinfo {author} {\bibfnamefont {Y.}~\bibnamefont
			{Sun}}, \bibinfo {author} {\bibfnamefont {S.-C.}\ \bibnamefont {Wu}},\ and\
		\bibinfo {author} {\bibfnamefont {B.}~\bibnamefont {Yan}},\ }\bibfield
	{title} {\bibinfo {title} {Topological surface states and $\mathrm{Fermi}$
			arcs of the noncentrosymmetric $\mathrm{Weyl}$ semimetals $\mathrm{TaAs}$,
			$\mathrm{TaP}$, $\mathrm{NbAs}$, and $\mathrm{NbP}$},\ }\href
	{https://doi.org/10.1103/PhysRevB.92.115428} {\bibfield  {journal} {\bibinfo
			{journal} {Phys. Rev. B}\ }\textbf {\bibinfo {volume} {92}},\ \bibinfo
		{pages} {115428} (\bibinfo {year} {2015})}\BibitemShut {NoStop}%
	\bibitem [{\citenamefont {Souma}\ \emph {et~al.}(2016)\citenamefont {Souma},
		\citenamefont {Wang}, \citenamefont {Kotaka}, \citenamefont {Sato},
		\citenamefont {Nakayama}, \citenamefont {Tanaka}, \citenamefont {Kimizuka},
		\citenamefont {Takahashi}, \citenamefont {Yamauchi}, \citenamefont {Oguchi},
		\citenamefont {Segawa},\ and\ \citenamefont {Ando}}]{PhysRevB.93.161112}%
	\BibitemOpen
	\bibfield  {author} {\bibinfo {author} {\bibfnamefont {S.}~\bibnamefont
			{Souma}}, \bibinfo {author} {\bibfnamefont {Z.}~\bibnamefont {Wang}},
		\bibinfo {author} {\bibfnamefont {H.}~\bibnamefont {Kotaka}}, \bibinfo
		{author} {\bibfnamefont {T.}~\bibnamefont {Sato}}, \bibinfo {author}
		{\bibfnamefont {K.}~\bibnamefont {Nakayama}}, \bibinfo {author}
		{\bibfnamefont {Y.}~\bibnamefont {Tanaka}}, \bibinfo {author} {\bibfnamefont
			{H.}~\bibnamefont {Kimizuka}}, \bibinfo {author} {\bibfnamefont
			{T.}~\bibnamefont {Takahashi}}, \bibinfo {author} {\bibfnamefont
			{K.}~\bibnamefont {Yamauchi}}, \bibinfo {author} {\bibfnamefont
			{T.}~\bibnamefont {Oguchi}}, \bibinfo {author} {\bibfnamefont
			{K.}~\bibnamefont {Segawa}},\ and\ \bibinfo {author} {\bibfnamefont
			{Y.}~\bibnamefont {Ando}},\ }\bibfield  {title} {\bibinfo {title} {Direct
			observation of nonequivalent $\mathrm{Fermi}$-arc states of opposite surfaces
			in the noncentrosymmetric $\mathrm{Weyl}$ semimetal $\mathrm{NbP}$},\ }\href
	{https://doi.org/10.1103/PhysRevB.93.161112} {\bibfield  {journal} {\bibinfo
			{journal} {Phys. Rev. B}\ }\textbf {\bibinfo {volume} {93}},\ \bibinfo
		{pages} {161112} (\bibinfo {year} {2016})}\BibitemShut {NoStop}%
	\bibitem [{\citenamefont {Potter}\ \emph {et~al.}(2014)\citenamefont {Potter},
		\citenamefont {Kimchi},\ and\ \citenamefont
		{Vishwanath}}]{potter2014quantum}%
	\BibitemOpen
	\bibfield  {author} {\bibinfo {author} {\bibfnamefont {A.~C.}\ \bibnamefont
			{Potter}}, \bibinfo {author} {\bibfnamefont {I.}~\bibnamefont {Kimchi}},\
		and\ \bibinfo {author} {\bibfnamefont {A.}~\bibnamefont {Vishwanath}},\
	}\bibfield  {title} {\bibinfo {title} {Quantum oscillations from surface
			$\mathrm{Fermi}$ arcs in $\mathrm{Weyl}$ and $\mathrm{Dirac}$ semimetals},\
	}\href {https://doi.org/10.1038/ncomms6161} {\bibfield  {journal} {\bibinfo
			{journal} {Nature Communications}\ }\textbf {\bibinfo {volume} {5}},\
		\bibinfo {pages} {5161} (\bibinfo {year} {2014})}\BibitemShut {NoStop}%
	\bibitem [{\citenamefont {Moll}\ \emph {et~al.}(2016)\citenamefont {Moll},
		\citenamefont {Nair}, \citenamefont {Helm}, \citenamefont {Potter},
		\citenamefont {Kimchi}, \citenamefont {Vishwanath},\ and\ \citenamefont
		{Analytis}}]{moll2016transport}%
	\BibitemOpen
	\bibfield  {author} {\bibinfo {author} {\bibfnamefont {P.~J.}\ \bibnamefont
			{Moll}}, \bibinfo {author} {\bibfnamefont {N.~L.}\ \bibnamefont {Nair}},
		\bibinfo {author} {\bibfnamefont {T.}~\bibnamefont {Helm}}, \bibinfo {author}
		{\bibfnamefont {A.~C.}\ \bibnamefont {Potter}}, \bibinfo {author}
		{\bibfnamefont {I.}~\bibnamefont {Kimchi}}, \bibinfo {author} {\bibfnamefont
			{A.}~\bibnamefont {Vishwanath}},\ and\ \bibinfo {author} {\bibfnamefont
			{J.~G.}\ \bibnamefont {Analytis}},\ }\bibfield  {title} {\bibinfo {title}
		{Transport evidence for $\mathrm{Fermi}$-arc-mediated chirality transfer in
			the $\mathrm{Dirac}$ semimetal $\mathrm{Cd}_{3}\mathrm{As}_{2}$},\ }\href
	{https://doi.org/10.1038/nature18276} {\bibfield  {journal} {\bibinfo
			{journal} {Nature}\ }\textbf {\bibinfo {volume} {535}},\ \bibinfo {pages}
		{266} (\bibinfo {year} {2016})}\BibitemShut {NoStop}%
	\bibitem [{\citenamefont {Morali}\ \emph {et~al.}(2019)\citenamefont {Morali},
		\citenamefont {Batabyal}, \citenamefont {Nag}, \citenamefont {Liu},
		\citenamefont {Xu}, \citenamefont {Sun}, \citenamefont {Yan}, \citenamefont
		{Felser}, \citenamefont {Avraham},\ and\ \citenamefont
		{Beidenkopf}}]{science.aav2334}%
	\BibitemOpen
	\bibfield  {author} {\bibinfo {author} {\bibfnamefont {N.}~\bibnamefont
			{Morali}}, \bibinfo {author} {\bibfnamefont {R.}~\bibnamefont {Batabyal}},
		\bibinfo {author} {\bibfnamefont {P.~K.}\ \bibnamefont {Nag}}, \bibinfo
		{author} {\bibfnamefont {E.}~\bibnamefont {Liu}}, \bibinfo {author}
		{\bibfnamefont {Q.}~\bibnamefont {Xu}}, \bibinfo {author} {\bibfnamefont
			{Y.}~\bibnamefont {Sun}}, \bibinfo {author} {\bibfnamefont {B.}~\bibnamefont
			{Yan}}, \bibinfo {author} {\bibfnamefont {C.}~\bibnamefont {Felser}},
		\bibinfo {author} {\bibfnamefont {N.}~\bibnamefont {Avraham}},\ and\ \bibinfo
		{author} {\bibfnamefont {H.}~\bibnamefont {Beidenkopf}},\ }\bibfield  {title}
	{\bibinfo {title} {Fermi-arc diversity on surface terminations of the
			magnetic $\mathrm{Weyl}$ semimetal
			$\mathrm{Co}_{3}\mathrm{Sn}_{2}\mathrm{S}_{2}$},\ }\href
	{https://doi.org/10.1126/science.aav2334} {\bibfield  {journal} {\bibinfo
			{journal} {Science}\ }\textbf {\bibinfo {volume} {365}},\ \bibinfo {pages}
		{1286} (\bibinfo {year} {2019})}\BibitemShut {NoStop}%
	\bibitem [{\citenamefont {Vaqueiro}\ and\ \citenamefont
		{Sobany}(2009)}]{VAQUEIRO2009513}%
	\BibitemOpen
	\bibfield  {author} {\bibinfo {author} {\bibfnamefont {P.}~\bibnamefont
			{Vaqueiro}}\ and\ \bibinfo {author} {\bibfnamefont {G.~G.}\ \bibnamefont
			{Sobany}},\ }\bibfield  {title} {\bibinfo {title} {A powder neutron
			diffraction study of the metallic ferromagnet
			$\mathrm{Co}_{3}\mathrm{Sn}_{2}\mathrm{S}_{2}$},\ }\href
	{https://doi.org/https://doi.org/10.1016/j.solidstatesciences.2008.06.017}
	{\bibfield  {journal} {\bibinfo  {journal} {Solid State Sciences}\ }\textbf
		{\bibinfo {volume} {11}},\ \bibinfo {pages} {513} (\bibinfo {year}
		{2009})}\BibitemShut {NoStop}%
	\bibitem [{\citenamefont {Xu}\ \emph {et~al.}(2020)\citenamefont {Xu},
		\citenamefont {Zhao}, \citenamefont {Yi}, \citenamefont {Wang}, \citenamefont
		{Yin}, \citenamefont {Wang}, \citenamefont {Hu}, \citenamefont {Wang},
		\citenamefont {Liu}, \citenamefont {Xu} \emph {et~al.}}]{xu2020electronic}%
	\BibitemOpen
	\bibfield  {author} {\bibinfo {author} {\bibfnamefont {Y.}~\bibnamefont
			{Xu}}, \bibinfo {author} {\bibfnamefont {J.}~\bibnamefont {Zhao}}, \bibinfo
		{author} {\bibfnamefont {C.}~\bibnamefont {Yi}}, \bibinfo {author}
		{\bibfnamefont {Q.}~\bibnamefont {Wang}}, \bibinfo {author} {\bibfnamefont
			{Q.}~\bibnamefont {Yin}}, \bibinfo {author} {\bibfnamefont {Y.}~\bibnamefont
			{Wang}}, \bibinfo {author} {\bibfnamefont {X.}~\bibnamefont {Hu}}, \bibinfo
		{author} {\bibfnamefont {L.}~\bibnamefont {Wang}}, \bibinfo {author}
		{\bibfnamefont {E.}~\bibnamefont {Liu}}, \bibinfo {author} {\bibfnamefont
			{G.}~\bibnamefont {Xu}}, \emph {et~al.},\ }\bibfield  {title} {\bibinfo
		{title} {Electronic correlations and flattened band in magnetic
			$\mathrm{Weyl}$ semimetal candidate
			$\mathrm{Co}_{3}\mathrm{Sn}_{2}\mathrm{S}_{2}$},\ }\href
	{https://doi.org/10.1038/s41467-020-17234-0} {\bibfield  {journal} {\bibinfo
			{journal} {Nature Communications}\ }\textbf {\bibinfo {volume} {11}},\
		\bibinfo {pages} {3985} (\bibinfo {year} {2020})}\BibitemShut {NoStop}%
	\bibitem [{\citenamefont {Weihrich}\ \emph {et~al.}(2005)\citenamefont
		{Weihrich}, \citenamefont {Anusca},\ and\ \citenamefont
		{Zabel}}]{zaac.200400561}%
	\BibitemOpen
	\bibfield  {author} {\bibinfo {author} {\bibfnamefont {R.}~\bibnamefont
			{Weihrich}}, \bibinfo {author} {\bibfnamefont {I.}~\bibnamefont {Anusca}},\
		and\ \bibinfo {author} {\bibfnamefont {M.}~\bibnamefont {Zabel}},\ }\bibfield
	{title} {\bibinfo {title} {Halbantiperowskite: $\mathrm{Zur}$
			$\mathrm{Struktur}$ der $\mathrm{Shandite}$ $\mathrm{M}_{3/2}\mathrm{As}$
			($\mathrm{M}=\mathrm{Co},\mathrm{Ni}$; $\mathrm{A}=\mathrm{In},\mathrm{Sn}$)
			und ihren $\mathrm{Typ}$-$\mathrm{Antitypbeziehungen}$},\ }\href
	{https://doi.org/https://doi.org/10.1002/zaac.200400561} {\bibfield
		{journal} {\bibinfo  {journal} {Zeitschrift Für Anorganische Und Allgemeine
				Chemie}\ }\textbf {\bibinfo {volume} {631}},\ \bibinfo {pages} {1463}
		(\bibinfo {year} {2005})}\BibitemShut {NoStop}%
	\bibitem [{\citenamefont {Yoshikawa}\ \emph {et~al.}(2022)\citenamefont
		{Yoshikawa}, \citenamefont {Ogawa}, \citenamefont {Hirai}, \citenamefont
		{Fujiwara}, \citenamefont {Ikeda}, \citenamefont {Tsukazaki},\ and\
		\citenamefont {Shimano}}]{yoshikawa2022non}%
	\BibitemOpen
	\bibfield  {author} {\bibinfo {author} {\bibfnamefont {N.}~\bibnamefont
			{Yoshikawa}}, \bibinfo {author} {\bibfnamefont {K.}~\bibnamefont {Ogawa}},
		\bibinfo {author} {\bibfnamefont {Y.}~\bibnamefont {Hirai}}, \bibinfo
		{author} {\bibfnamefont {K.}~\bibnamefont {Fujiwara}}, \bibinfo {author}
		{\bibfnamefont {J.}~\bibnamefont {Ikeda}}, \bibinfo {author} {\bibfnamefont
			{A.}~\bibnamefont {Tsukazaki}},\ and\ \bibinfo {author} {\bibfnamefont
			{R.}~\bibnamefont {Shimano}},\ }\bibfield  {title} {\bibinfo {title}
		{Non-volatile chirality switching by all-optical magnetization reversal in
			ferromagnetic $\mathrm{Weyl}$ semimetal
			$\mathrm{Co}_{3}\mathrm{Sn}_{2}\mathrm{S}_{2}$},\ }\href
	{https://doi.org/10.1038/s42005-022-01106-8} {\bibfield  {journal} {\bibinfo
			{journal} {Communications Physics}\ }\textbf {\bibinfo {volume} {5}},\
		\bibinfo {pages} {328} (\bibinfo {year} {2022})}\BibitemShut {NoStop}%
	\bibitem [{\citenamefont {Ikeda}\ \emph
		{et~al.}(2021{\natexlab{a}})\citenamefont {Ikeda}, \citenamefont {Fujiwara},
		\citenamefont {Shiogai}, \citenamefont {Seki}, \citenamefont {Nomura},
		\citenamefont {Takanashi},\ and\ \citenamefont
		{Tsukazaki}}]{ikeda2021critical}%
	\BibitemOpen
	\bibfield  {author} {\bibinfo {author} {\bibfnamefont {J.}~\bibnamefont
			{Ikeda}}, \bibinfo {author} {\bibfnamefont {K.}~\bibnamefont {Fujiwara}},
		\bibinfo {author} {\bibfnamefont {J.}~\bibnamefont {Shiogai}}, \bibinfo
		{author} {\bibfnamefont {T.}~\bibnamefont {Seki}}, \bibinfo {author}
		{\bibfnamefont {K.}~\bibnamefont {Nomura}}, \bibinfo {author} {\bibfnamefont
			{K.}~\bibnamefont {Takanashi}},\ and\ \bibinfo {author} {\bibfnamefont
			{A.}~\bibnamefont {Tsukazaki}},\ }\bibfield  {title} {\bibinfo {title}
		{Critical thickness for the emergence of $\mathrm{Weyl}$ features in
			$\mathrm{Co}_{3}\mathrm{Sn}_{2}\mathrm{S}_{2}$ thin films},\ }\href
	{https://doi.org/10.1038/s43246-021-00122-5} {\bibfield  {journal} {\bibinfo
			{journal} {Communications Materials}\ }\textbf {\bibinfo {volume} {2}},\
		\bibinfo {pages} {18} (\bibinfo {year} {2021}{\natexlab{a}})}\BibitemShut
	{NoStop}%
	\bibitem [{\citenamefont {Wang}\ \emph {et~al.}(2018)\citenamefont {Wang},
		\citenamefont {Xu}, \citenamefont {Lou}, \citenamefont {Liu}, \citenamefont
		{Li}, \citenamefont {Huang}, \citenamefont {Shen}, \citenamefont {Weng},
		\citenamefont {Wang},\ and\ \citenamefont {Lei}}]{wang2018large}%
	\BibitemOpen
	\bibfield  {author} {\bibinfo {author} {\bibfnamefont {Q.}~\bibnamefont
			{Wang}}, \bibinfo {author} {\bibfnamefont {Y.}~\bibnamefont {Xu}}, \bibinfo
		{author} {\bibfnamefont {R.}~\bibnamefont {Lou}}, \bibinfo {author}
		{\bibfnamefont {Z.}~\bibnamefont {Liu}}, \bibinfo {author} {\bibfnamefont
			{M.}~\bibnamefont {Li}}, \bibinfo {author} {\bibfnamefont {Y.}~\bibnamefont
			{Huang}}, \bibinfo {author} {\bibfnamefont {D.}~\bibnamefont {Shen}},
		\bibinfo {author} {\bibfnamefont {H.}~\bibnamefont {Weng}}, \bibinfo {author}
		{\bibfnamefont {S.}~\bibnamefont {Wang}},\ and\ \bibinfo {author}
		{\bibfnamefont {H.}~\bibnamefont {Lei}},\ }\bibfield  {title} {\bibinfo
		{title} {Large intrinsic anomalous $\mathrm{Hall}$ effect in half-metallic
			ferromagnet $\mathrm{Co}_{3}\mathrm{Sn}_{2}\mathrm{S}_{2}$ with magnetic
			$\mathrm{Weyl}$ fermions},\ }\href
	{https://doi.org/10.1038/s41467-018-06088-2} {\bibfield  {journal} {\bibinfo
			{journal} {Nature Communications}\ }\textbf {\bibinfo {volume} {9}},\
		\bibinfo {pages} {1} (\bibinfo {year} {2018})}\BibitemShut {NoStop}%
	\bibitem [{\citenamefont {Okamura}\ \emph {et~al.}(2020)\citenamefont
		{Okamura}, \citenamefont {Minami}, \citenamefont {Kato}, \citenamefont
		{Fujishiro}, \citenamefont {Kaneko}, \citenamefont {Ikeda}, \citenamefont
		{Muramoto}, \citenamefont {Kaneko}, \citenamefont {Ueda}, \citenamefont
		{Kocsis} \emph {et~al.}}]{okamura2020giant}%
	\BibitemOpen
	\bibfield  {author} {\bibinfo {author} {\bibfnamefont {Y.}~\bibnamefont
			{Okamura}}, \bibinfo {author} {\bibfnamefont {S.}~\bibnamefont {Minami}},
		\bibinfo {author} {\bibfnamefont {Y.}~\bibnamefont {Kato}}, \bibinfo {author}
		{\bibfnamefont {Y.}~\bibnamefont {Fujishiro}}, \bibinfo {author}
		{\bibfnamefont {Y.}~\bibnamefont {Kaneko}}, \bibinfo {author} {\bibfnamefont
			{J.}~\bibnamefont {Ikeda}}, \bibinfo {author} {\bibfnamefont
			{J.}~\bibnamefont {Muramoto}}, \bibinfo {author} {\bibfnamefont
			{R.}~\bibnamefont {Kaneko}}, \bibinfo {author} {\bibfnamefont
			{K.}~\bibnamefont {Ueda}}, \bibinfo {author} {\bibfnamefont {V.}~\bibnamefont
			{Kocsis}}, \emph {et~al.},\ }\bibfield  {title} {\bibinfo {title} {Giant
			magneto-optical responses in magnetic $\mathrm{Weyl}$ semimetal
			$\mathrm{Co}_{3}\mathrm{Sn}_{2}\mathrm{S}_{2}$},\ }\href
	{https://doi.org/10.1038/s41467-020-18470-0} {\bibfield  {journal} {\bibinfo
			{journal} {Nature Communications}\ }\textbf {\bibinfo {volume} {11}},\
		\bibinfo {pages} {4619} (\bibinfo {year} {2020})}\BibitemShut {NoStop}%
	\bibitem [{\citenamefont {Ding}\ \emph {et~al.}(2019)\citenamefont {Ding},
		\citenamefont {Koo}, \citenamefont {Xu}, \citenamefont {Li}, \citenamefont
		{Lu}, \citenamefont {Zhao}, \citenamefont {Wang}, \citenamefont {Yin},
		\citenamefont {Lei}, \citenamefont {Yan}, \citenamefont {Zhu},\ and\
		\citenamefont {Behnia}}]{PhysRevX.9.041061}%
	\BibitemOpen
	\bibfield  {author} {\bibinfo {author} {\bibfnamefont {L.}~\bibnamefont
			{Ding}}, \bibinfo {author} {\bibfnamefont {J.}~\bibnamefont {Koo}}, \bibinfo
		{author} {\bibfnamefont {L.}~\bibnamefont {Xu}}, \bibinfo {author}
		{\bibfnamefont {X.}~\bibnamefont {Li}}, \bibinfo {author} {\bibfnamefont
			{X.}~\bibnamefont {Lu}}, \bibinfo {author} {\bibfnamefont {L.}~\bibnamefont
			{Zhao}}, \bibinfo {author} {\bibfnamefont {Q.}~\bibnamefont {Wang}}, \bibinfo
		{author} {\bibfnamefont {Q.}~\bibnamefont {Yin}}, \bibinfo {author}
		{\bibfnamefont {H.}~\bibnamefont {Lei}}, \bibinfo {author} {\bibfnamefont
			{B.}~\bibnamefont {Yan}}, \bibinfo {author} {\bibfnamefont {Z.}~\bibnamefont
			{Zhu}},\ and\ \bibinfo {author} {\bibfnamefont {K.}~\bibnamefont {Behnia}},\
	}\bibfield  {title} {\bibinfo {title} {Intrinsic $\mathrm{Anomalous}$
			$\mathrm{Nernst}$ $\mathrm{Effect}$ $\mathrm{Amplified}$ by
			$\mathrm{Disorder}$ in a $\mathrm{Half}$-$\mathrm{Metallic}$
			$\mathrm{Semimetal}$},\ }\href {https://doi.org/10.1103/PhysRevX.9.041061}
	{\bibfield  {journal} {\bibinfo  {journal} {Phys. Rev. X}\ }\textbf {\bibinfo
			{volume} {9}},\ \bibinfo {pages} {041061} (\bibinfo {year}
		{2019})}\BibitemShut {NoStop}%
	\bibitem [{\citenamefont {Guin}\ \emph {et~al.}(2019)\citenamefont {Guin},
		\citenamefont {Vir}, \citenamefont {Zhang}, \citenamefont {Kumar},
		\citenamefont {Watzman}, \citenamefont {Fu}, \citenamefont {Liu},
		\citenamefont {Manna}, \citenamefont {Schnelle}, \citenamefont {Gooth},
		\citenamefont {Shekhar}, \citenamefont {Sun},\ and\ \citenamefont
		{Felser}}]{adma.201806622}%
	\BibitemOpen
	\bibfield  {author} {\bibinfo {author} {\bibfnamefont {S.~N.}\ \bibnamefont
			{Guin}}, \bibinfo {author} {\bibfnamefont {P.}~\bibnamefont {Vir}}, \bibinfo
		{author} {\bibfnamefont {Y.}~\bibnamefont {Zhang}}, \bibinfo {author}
		{\bibfnamefont {N.}~\bibnamefont {Kumar}}, \bibinfo {author} {\bibfnamefont
			{S.~J.}\ \bibnamefont {Watzman}}, \bibinfo {author} {\bibfnamefont
			{C.}~\bibnamefont {Fu}}, \bibinfo {author} {\bibfnamefont {E.}~\bibnamefont
			{Liu}}, \bibinfo {author} {\bibfnamefont {K.}~\bibnamefont {Manna}}, \bibinfo
		{author} {\bibfnamefont {W.}~\bibnamefont {Schnelle}}, \bibinfo {author}
		{\bibfnamefont {J.}~\bibnamefont {Gooth}}, \bibinfo {author} {\bibfnamefont
			{C.}~\bibnamefont {Shekhar}}, \bibinfo {author} {\bibfnamefont
			{Y.}~\bibnamefont {Sun}},\ and\ \bibinfo {author} {\bibfnamefont
			{C.}~\bibnamefont {Felser}},\ }\bibfield  {title} {\bibinfo {title}
		{Zero-$\mathrm{Field}$ $\mathrm{Nernst}$ $\mathrm{Effect}$ in a
			$\mathrm{Ferromagnetic}$ $\mathrm{Kagome}$-$\mathrm{Lattice}$
			$\mathrm{Weyl}$-$\mathrm{Semimetal}$
			$\mathrm{Co}_{3}\mathrm{Sn}_{2}\mathrm{S}_{2}$},\ }\href
	{https://doi.org/https://doi.org/10.1002/adma.201806622} {\bibfield
		{journal} {\bibinfo  {journal} {Advanced Materials}\ }\textbf {\bibinfo
			{volume} {31}},\ \bibinfo {pages} {1806622} (\bibinfo {year}
		{2019})}\BibitemShut {NoStop}%
	\bibitem [{\citenamefont {Yang}\ \emph
		{et~al.}(2020{\natexlab{a}})\citenamefont {Yang}, \citenamefont {You},
		\citenamefont {Wang}, \citenamefont {Huang}, \citenamefont {Xi},
		\citenamefont {Xu}, \citenamefont {Cao}, \citenamefont {Tian}, \citenamefont
		{Xu}, \citenamefont {Dai},\ and\ \citenamefont
		{Li}}]{PhysRevMaterials.4.024202}%
	\BibitemOpen
	\bibfield  {author} {\bibinfo {author} {\bibfnamefont {H.}~\bibnamefont
			{Yang}}, \bibinfo {author} {\bibfnamefont {W.}~\bibnamefont {You}}, \bibinfo
		{author} {\bibfnamefont {J.}~\bibnamefont {Wang}}, \bibinfo {author}
		{\bibfnamefont {J.}~\bibnamefont {Huang}}, \bibinfo {author} {\bibfnamefont
			{C.}~\bibnamefont {Xi}}, \bibinfo {author} {\bibfnamefont {X.}~\bibnamefont
			{Xu}}, \bibinfo {author} {\bibfnamefont {C.}~\bibnamefont {Cao}}, \bibinfo
		{author} {\bibfnamefont {M.}~\bibnamefont {Tian}}, \bibinfo {author}
		{\bibfnamefont {Z.-A.}\ \bibnamefont {Xu}}, \bibinfo {author} {\bibfnamefont
			{J.}~\bibnamefont {Dai}},\ and\ \bibinfo {author} {\bibfnamefont
			{Y.}~\bibnamefont {Li}},\ }\bibfield  {title} {\bibinfo {title} {Giant
			anomalous $\mathrm{Nernst}$ effect in the magnetic $\mathrm{Weyl}$ semimetal
			$\mathrm{Co}_{3}\mathrm{Sn}_{2}\mathrm{S}_{2}$},\ }\href
	{https://doi.org/10.1103/PhysRevMaterials.4.024202} {\bibfield  {journal}
		{\bibinfo  {journal} {Phys. Rev. Mater.}\ }\textbf {\bibinfo {volume} {4}},\
		\bibinfo {pages} {024202} (\bibinfo {year} {2020}{\natexlab{a}})}\BibitemShut
	{NoStop}%
	\bibitem [{\citenamefont {Liu}\ \emph {et~al.}(2019)\citenamefont {Liu},
		\citenamefont {Liang}, \citenamefont {Liu}, \citenamefont {Xu}, \citenamefont
		{Li}, \citenamefont {Chen}, \citenamefont {Pei}, \citenamefont {Shi},
		\citenamefont {Mo}, \citenamefont {Dudin}, \citenamefont {Kim}, \citenamefont
		{Cacho}, \citenamefont {Li}, \citenamefont {Sun}, \citenamefont {Yang},
		\citenamefont {Liu}, \citenamefont {Parkin}, \citenamefont {Felser},\ and\
		\citenamefont {Chen}}]{science.aav2873}%
	\BibitemOpen
	\bibfield  {author} {\bibinfo {author} {\bibfnamefont {D.~F.}\ \bibnamefont
			{Liu}}, \bibinfo {author} {\bibfnamefont {A.~J.}\ \bibnamefont {Liang}},
		\bibinfo {author} {\bibfnamefont {E.~K.}\ \bibnamefont {Liu}}, \bibinfo
		{author} {\bibfnamefont {Q.~N.}\ \bibnamefont {Xu}}, \bibinfo {author}
		{\bibfnamefont {Y.~W.}\ \bibnamefont {Li}}, \bibinfo {author} {\bibfnamefont
			{C.}~\bibnamefont {Chen}}, \bibinfo {author} {\bibfnamefont {D.}~\bibnamefont
			{Pei}}, \bibinfo {author} {\bibfnamefont {W.~J.}\ \bibnamefont {Shi}},
		\bibinfo {author} {\bibfnamefont {S.~K.}\ \bibnamefont {Mo}}, \bibinfo
		{author} {\bibfnamefont {P.}~\bibnamefont {Dudin}}, \bibinfo {author}
		{\bibfnamefont {T.}~\bibnamefont {Kim}}, \bibinfo {author} {\bibfnamefont
			{C.}~\bibnamefont {Cacho}}, \bibinfo {author} {\bibfnamefont
			{G.}~\bibnamefont {Li}}, \bibinfo {author} {\bibfnamefont {Y.}~\bibnamefont
			{Sun}}, \bibinfo {author} {\bibfnamefont {L.~X.}\ \bibnamefont {Yang}},
		\bibinfo {author} {\bibfnamefont {Z.~K.}\ \bibnamefont {Liu}}, \bibinfo
		{author} {\bibfnamefont {S.~S.~P.}\ \bibnamefont {Parkin}}, \bibinfo {author}
		{\bibfnamefont {C.}~\bibnamefont {Felser}},\ and\ \bibinfo {author}
		{\bibfnamefont {Y.~L.}\ \bibnamefont {Chen}},\ }\bibfield  {title} {\bibinfo
		{title} {Magnetic $\mathrm{Weyl}$ semimetal phase in a $\mathrm{Kagom}$é
			crystal},\ }\href {https://doi.org/10.1126/science.aav2873} {\bibfield
		{journal} {\bibinfo  {journal} {Science}\ }\textbf {\bibinfo {volume}
			{365}},\ \bibinfo {pages} {1282} (\bibinfo {year} {2019})}\BibitemShut
	{NoStop}%
	\bibitem [{\citenamefont {Ikeda}\ \emph
		{et~al.}(2021{\natexlab{b}})\citenamefont {Ikeda}, \citenamefont {Fujiwara},
		\citenamefont {Shiogai}, \citenamefont {Seki}, \citenamefont {Nomura},
		\citenamefont {Takanashi},\ and\ \citenamefont {Tsukazaki}}]{ikeda2021two}%
	\BibitemOpen
	\bibfield  {author} {\bibinfo {author} {\bibfnamefont {J.}~\bibnamefont
			{Ikeda}}, \bibinfo {author} {\bibfnamefont {K.}~\bibnamefont {Fujiwara}},
		\bibinfo {author} {\bibfnamefont {J.}~\bibnamefont {Shiogai}}, \bibinfo
		{author} {\bibfnamefont {T.}~\bibnamefont {Seki}}, \bibinfo {author}
		{\bibfnamefont {K.}~\bibnamefont {Nomura}}, \bibinfo {author} {\bibfnamefont
			{K.}~\bibnamefont {Takanashi}},\ and\ \bibinfo {author} {\bibfnamefont
			{A.}~\bibnamefont {Tsukazaki}},\ }\bibfield  {title} {\bibinfo {title}
		{Two-dimensionality of metallic surface conduction in
			$\mathrm{Co}_{3}\mathrm{Sn}_{2}\mathrm{S}_{2}$ thin films},\ }\href
	{https://doi.org/10.1038/s42005-021-00627-y} {\bibfield  {journal} {\bibinfo
			{journal} {Communications Physics}\ }\textbf {\bibinfo {volume} {4}},\
		\bibinfo {pages} {117} (\bibinfo {year} {2021}{\natexlab{b}})}\BibitemShut
	{NoStop}%
	\bibitem [{\citenamefont {Howard}\ \emph {et~al.}(2021)\citenamefont {Howard},
		\citenamefont {Jiao}, \citenamefont {Wang}, \citenamefont {Morali},
		\citenamefont {Batabyal}, \citenamefont {Kumar-Nag}, \citenamefont {Avraham},
		\citenamefont {Beidenkopf}, \citenamefont {Vir}, \citenamefont {Liu} \emph
		{et~al.}}]{howard2021evidence}%
	\BibitemOpen
	\bibfield  {author} {\bibinfo {author} {\bibfnamefont {S.}~\bibnamefont
			{Howard}}, \bibinfo {author} {\bibfnamefont {L.}~\bibnamefont {Jiao}},
		\bibinfo {author} {\bibfnamefont {Z.}~\bibnamefont {Wang}}, \bibinfo {author}
		{\bibfnamefont {N.}~\bibnamefont {Morali}}, \bibinfo {author} {\bibfnamefont
			{R.}~\bibnamefont {Batabyal}}, \bibinfo {author} {\bibfnamefont
			{P.}~\bibnamefont {Kumar-Nag}}, \bibinfo {author} {\bibfnamefont
			{N.}~\bibnamefont {Avraham}}, \bibinfo {author} {\bibfnamefont
			{H.}~\bibnamefont {Beidenkopf}}, \bibinfo {author} {\bibfnamefont
			{P.}~\bibnamefont {Vir}}, \bibinfo {author} {\bibfnamefont {E.}~\bibnamefont
			{Liu}}, \emph {et~al.},\ }\bibfield  {title} {\bibinfo {title} {Evidence for
			one-dimensional chiral edge states in a magnetic $\mathrm{Weyl}$ semimetal
			$\mathrm{Co}_{3}\mathrm{Sn}_{2}\mathrm{S}_{2}$},\ }\href
	{https://doi.org/10.1038/s41467-021-24561-3} {\bibfield  {journal} {\bibinfo
			{journal} {Nature Communications}\ }\textbf {\bibinfo {volume} {12}},\
		\bibinfo {pages} {4269} (\bibinfo {year} {2021})}\BibitemShut {NoStop}%
	\bibitem [{\citenamefont {Xu}\ \emph {et~al.}(2018)\citenamefont {Xu},
		\citenamefont {Liu}, \citenamefont {Shi}, \citenamefont {Muechler},
		\citenamefont {Gayles}, \citenamefont {Felser},\ and\ \citenamefont
		{Sun}}]{PhysRevB.97.235416}%
	\BibitemOpen
	\bibfield  {author} {\bibinfo {author} {\bibfnamefont {Q.}~\bibnamefont
			{Xu}}, \bibinfo {author} {\bibfnamefont {E.}~\bibnamefont {Liu}}, \bibinfo
		{author} {\bibfnamefont {W.}~\bibnamefont {Shi}}, \bibinfo {author}
		{\bibfnamefont {L.}~\bibnamefont {Muechler}}, \bibinfo {author}
		{\bibfnamefont {J.}~\bibnamefont {Gayles}}, \bibinfo {author} {\bibfnamefont
			{C.}~\bibnamefont {Felser}},\ and\ \bibinfo {author} {\bibfnamefont
			{Y.}~\bibnamefont {Sun}},\ }\bibfield  {title} {\bibinfo {title} {Topological
			surface $\mathrm{Fermi}$ arcs in the magnetic $\mathrm{Weyl}$ semimetal
			$\mathrm{Co}_{3}\mathrm{Sn}_{2}\mathrm{S}_{2}$},\ }\href
	{https://doi.org/10.1103/PhysRevB.97.235416} {\bibfield  {journal} {\bibinfo
			{journal} {Phys. Rev. B}\ }\textbf {\bibinfo {volume} {97}},\ \bibinfo
		{pages} {235416} (\bibinfo {year} {2018})}\BibitemShut {NoStop}%
	\bibitem [{\citenamefont {Bernevig}\ \emph {et~al.}(2022)\citenamefont
		{Bernevig}, \citenamefont {Felser},\ and\ \citenamefont
		{Beidenkopf}}]{bernevig2022progress}%
	\BibitemOpen
	\bibfield  {author} {\bibinfo {author} {\bibfnamefont {B.~A.}\ \bibnamefont
			{Bernevig}}, \bibinfo {author} {\bibfnamefont {C.}~\bibnamefont {Felser}},\
		and\ \bibinfo {author} {\bibfnamefont {H.}~\bibnamefont {Beidenkopf}},\
	}\bibfield  {title} {\bibinfo {title} {Progress and prospects in magnetic
			topological materials},\ }\href {https://doi.org/10.1038/s41586-021-04105-x}
	{\bibfield  {journal} {\bibinfo  {journal} {Nature}\ }\textbf {\bibinfo
			{volume} {603}},\ \bibinfo {pages} {41} (\bibinfo {year} {2022})}\BibitemShut
	{NoStop}%
	\bibitem [{\citenamefont {Nag}\ \emph {et~al.}(2022)\citenamefont {Nag},
		\citenamefont {Peng}, \citenamefont {Li}, \citenamefont {Agrestini},
		\citenamefont {Robarts}, \citenamefont {Garc{\'\i}a-Fern{\'a}ndez},
		\citenamefont {Walters}, \citenamefont {Wang}, \citenamefont {Yin},
		\citenamefont {Lei} \emph {et~al.}}]{nag2022correlation}%
	\BibitemOpen
	\bibfield  {author} {\bibinfo {author} {\bibfnamefont {A.}~\bibnamefont
			{Nag}}, \bibinfo {author} {\bibfnamefont {Y.}~\bibnamefont {Peng}}, \bibinfo
		{author} {\bibfnamefont {J.}~\bibnamefont {Li}}, \bibinfo {author}
		{\bibfnamefont {S.}~\bibnamefont {Agrestini}}, \bibinfo {author}
		{\bibfnamefont {H.}~\bibnamefont {Robarts}}, \bibinfo {author} {\bibfnamefont
			{M.}~\bibnamefont {Garc{\'\i}a-Fern{\'a}ndez}}, \bibinfo {author}
		{\bibfnamefont {A.}~\bibnamefont {Walters}}, \bibinfo {author} {\bibfnamefont
			{Q.}~\bibnamefont {Wang}}, \bibinfo {author} {\bibfnamefont {Q.}~\bibnamefont
			{Yin}}, \bibinfo {author} {\bibfnamefont {H.}~\bibnamefont {Lei}}, \emph
		{et~al.},\ }\bibfield  {title} {\bibinfo {title} {Correlation driven
			near-flat band $\mathrm{Stoner}$ excitations in a $\mathrm{Kagome}$ magnet},\
	}\href {https://doi.org/10.1038/s41467-022-34933-y} {\bibfield  {journal}
		{\bibinfo  {journal} {Nature Communications}\ }\textbf {\bibinfo {volume}
			{13}},\ \bibinfo {pages} {7317} (\bibinfo {year} {2022})}\BibitemShut
	{NoStop}%
	\bibitem [{\citenamefont {Wei}\ \emph {et~al.}(2012)\citenamefont {Wei},
		\citenamefont {Chao},\ and\ \citenamefont {Aji}}]{PhysRevLett.109.196403}%
	\BibitemOpen
	\bibfield  {author} {\bibinfo {author} {\bibfnamefont {H.}~\bibnamefont
			{Wei}}, \bibinfo {author} {\bibfnamefont {S.-P.}\ \bibnamefont {Chao}},\ and\
		\bibinfo {author} {\bibfnamefont {V.}~\bibnamefont {Aji}},\ }\bibfield
	{title} {\bibinfo {title} {Excitonic $\mathrm{Phases}$ from $\mathrm{Weyl}$
			$\mathrm{Semimetals}$},\ }\href
	{https://doi.org/10.1103/PhysRevLett.109.196403} {\bibfield  {journal}
		{\bibinfo  {journal} {Phys. Rev. Lett.}\ }\textbf {\bibinfo {volume} {109}},\
		\bibinfo {pages} {196403} (\bibinfo {year} {2012})}\BibitemShut {NoStop}%
	\bibitem [{\citenamefont {Go}\ \emph {et~al.}(2012)\citenamefont {Go},
		\citenamefont {Witczak-Krempa}, \citenamefont {Jeon}, \citenamefont {Park},\
		and\ \citenamefont {Kim}}]{PhysRevLett.109.066401}%
	\BibitemOpen
	\bibfield  {author} {\bibinfo {author} {\bibfnamefont {A.}~\bibnamefont
			{Go}}, \bibinfo {author} {\bibfnamefont {W.}~\bibnamefont {Witczak-Krempa}},
		\bibinfo {author} {\bibfnamefont {G.~S.}\ \bibnamefont {Jeon}}, \bibinfo
		{author} {\bibfnamefont {K.}~\bibnamefont {Park}},\ and\ \bibinfo {author}
		{\bibfnamefont {Y.~B.}\ \bibnamefont {Kim}},\ }\bibfield  {title} {\bibinfo
		{title} {Correlation $\mathrm{Effects}$ on $\mathrm{3D}$
			$\mathrm{Topological}$ $\mathrm{Phases}$: $\mathrm{From}$ $\mathrm{Bulk}$ to
			$\mathrm{Boundary}$},\ }\href
	{https://doi.org/10.1103/PhysRevLett.109.066401} {\bibfield  {journal}
		{\bibinfo  {journal} {Phys. Rev. Lett.}\ }\textbf {\bibinfo {volume} {109}},\
		\bibinfo {pages} {066401} (\bibinfo {year} {2012})}\BibitemShut {NoStop}%
	\bibitem [{\citenamefont {Wang}\ and\ \citenamefont
		{Zhang}(2013)}]{PhysRevB.87.161107}%
	\BibitemOpen
	\bibfield  {author} {\bibinfo {author} {\bibfnamefont {Z.}~\bibnamefont
			{Wang}}\ and\ \bibinfo {author} {\bibfnamefont {S.-C.}\ \bibnamefont
			{Zhang}},\ }\bibfield  {title} {\bibinfo {title} {Chiral anomaly, charge
			density waves, and axion strings from $\mathrm{Weyl}$ semimetals},\ }\href
	{https://doi.org/10.1103/PhysRevB.87.161107} {\bibfield  {journal} {\bibinfo
			{journal} {Phys. Rev. B}\ }\textbf {\bibinfo {volume} {87}},\ \bibinfo
		{pages} {161107} (\bibinfo {year} {2013})}\BibitemShut {NoStop}%
	\bibitem [{\citenamefont {Witczak-Krempa}\ \emph {et~al.}(2014)\citenamefont
		{Witczak-Krempa}, \citenamefont {Knap},\ and\ \citenamefont
		{Abanin}}]{PhysRevLett.113.136402}%
	\BibitemOpen
	\bibfield  {author} {\bibinfo {author} {\bibfnamefont {W.}~\bibnamefont
			{Witczak-Krempa}}, \bibinfo {author} {\bibfnamefont {M.}~\bibnamefont
			{Knap}},\ and\ \bibinfo {author} {\bibfnamefont {D.}~\bibnamefont {Abanin}},\
	}\bibfield  {title} {\bibinfo {title} {Interacting $\mathrm{Weyl}$
			$\mathrm{Semimetals}$: $\mathrm{Characterization}$ via the
			$\mathrm{Topological}$ $\mathrm{Hamiltonian}$ and its $\mathrm{Breakdown}$},\
	}\href {https://doi.org/10.1103/PhysRevLett.113.136402} {\bibfield  {journal}
		{\bibinfo  {journal} {Phys. Rev. Lett.}\ }\textbf {\bibinfo {volume} {113}},\
		\bibinfo {pages} {136402} (\bibinfo {year} {2014})}\BibitemShut {NoStop}%
	\bibitem [{\citenamefont {Sekine}\ and\ \citenamefont
		{Nomura}(2014)}]{JPSJ.83.094710}%
	\BibitemOpen
	\bibfield  {author} {\bibinfo {author} {\bibfnamefont {A.}~\bibnamefont
			{Sekine}}\ and\ \bibinfo {author} {\bibfnamefont {K.}~\bibnamefont
			{Nomura}},\ }\bibfield  {title} {\bibinfo {title} {Weyl $\mathrm{Semimetal}$
			in the $\mathrm{Strong}$ $\mathrm{Coulomb}$ $\mathrm{Interaction}$
			$\mathrm{Limit}$},\ }\href {https://doi.org/10.7566/JPSJ.83.094710}
	{\bibfield  {journal} {\bibinfo  {journal} {Journal of the Physical Society
				of Japan}\ }\textbf {\bibinfo {volume} {83}},\ \bibinfo {pages} {094710}
		(\bibinfo {year} {2014})}\BibitemShut {NoStop}%
	\bibitem [{\citenamefont {Bi}\ and\ \citenamefont
		{Wang}(2015)}]{PhysRevB.92.241109}%
	\BibitemOpen
	\bibfield  {author} {\bibinfo {author} {\bibfnamefont {R.}~\bibnamefont
			{Bi}}\ and\ \bibinfo {author} {\bibfnamefont {Z.}~\bibnamefont {Wang}},\
	}\bibfield  {title} {\bibinfo {title} {Unidirectional transport in electronic
			and photonic $\mathrm{Weyl}$ materials by $\mathrm{Dirac}$ mass
			engineering},\ }\href {https://doi.org/10.1103/PhysRevB.92.241109} {\bibfield
		{journal} {\bibinfo  {journal} {Phys. Rev. B}\ }\textbf {\bibinfo {volume}
			{92}},\ \bibinfo {pages} {241109} (\bibinfo {year} {2015})}\BibitemShut
	{NoStop}%
	\bibitem [{\citenamefont {Jian}\ \emph {et~al.}(2015)\citenamefont {Jian},
		\citenamefont {Jiang},\ and\ \citenamefont {Yao}}]{PhysRevLett.114.237001}%
	\BibitemOpen
	\bibfield  {author} {\bibinfo {author} {\bibfnamefont {S.-K.}\ \bibnamefont
			{Jian}}, \bibinfo {author} {\bibfnamefont {Y.-F.}\ \bibnamefont {Jiang}},\
		and\ \bibinfo {author} {\bibfnamefont {H.}~\bibnamefont {Yao}},\ }\bibfield
	{title} {\bibinfo {title} {Emergent $\mathrm{Spacetime}$
			$\mathrm{Supersymmetry}$ in $\mathrm{3D}$ $\mathrm{Weyl}$
			$\mathrm{Semimetals}$ and $\mathrm{2D}$ $\mathrm{Dirac}$
			$\mathrm{Semimetals}$},\ }\href
	{https://doi.org/10.1103/PhysRevLett.114.237001} {\bibfield  {journal}
		{\bibinfo  {journal} {Phys. Rev. Lett.}\ }\textbf {\bibinfo {volume} {114}},\
		\bibinfo {pages} {237001} (\bibinfo {year} {2015})}\BibitemShut {NoStop}%
	\bibitem [{\citenamefont {Morimoto}\ and\ \citenamefont
		{Nagaosa}(2016)}]{morimoto2016weyl}%
	\BibitemOpen
	\bibfield  {author} {\bibinfo {author} {\bibfnamefont {T.}~\bibnamefont
			{Morimoto}}\ and\ \bibinfo {author} {\bibfnamefont {N.}~\bibnamefont
			{Nagaosa}},\ }\bibfield  {title} {\bibinfo {title} {Weyl mott insulator},\
	}\href {https://doi.org/10.1038/srep19853} {\bibfield  {journal} {\bibinfo
			{journal} {Scientific Reports}\ }\textbf {\bibinfo {volume} {6}},\ \bibinfo
		{pages} {19853} (\bibinfo {year} {2016})}\BibitemShut {NoStop}%
	\bibitem [{\citenamefont {Wang}\ and\ \citenamefont
		{Ye}(2016)}]{PhysRevB.94.075115}%
	\BibitemOpen
	\bibfield  {author} {\bibinfo {author} {\bibfnamefont {Y.}~\bibnamefont
			{Wang}}\ and\ \bibinfo {author} {\bibfnamefont {P.}~\bibnamefont {Ye}},\
	}\bibfield  {title} {\bibinfo {title} {Topological density-wave states in a
			particle-hole symmetric $\mathrm{Weyl}$ metal},\ }\href
	{https://doi.org/10.1103/PhysRevB.94.075115} {\bibfield  {journal} {\bibinfo
			{journal} {Phys. Rev. B}\ }\textbf {\bibinfo {volume} {94}},\ \bibinfo
		{pages} {075115} (\bibinfo {year} {2016})}\BibitemShut {NoStop}%
	\bibitem [{\citenamefont {Laubach}\ \emph {et~al.}(2016)\citenamefont
		{Laubach}, \citenamefont {Platt}, \citenamefont {Thomale}, \citenamefont
		{Neupert},\ and\ \citenamefont {Rachel}}]{PhysRevB.94.241102}%
	\BibitemOpen
	\bibfield  {author} {\bibinfo {author} {\bibfnamefont {M.}~\bibnamefont
			{Laubach}}, \bibinfo {author} {\bibfnamefont {C.}~\bibnamefont {Platt}},
		\bibinfo {author} {\bibfnamefont {R.}~\bibnamefont {Thomale}}, \bibinfo
		{author} {\bibfnamefont {T.}~\bibnamefont {Neupert}},\ and\ \bibinfo {author}
		{\bibfnamefont {S.}~\bibnamefont {Rachel}},\ }\bibfield  {title} {\bibinfo
		{title} {Density wave instabilities and surface state evolution in
			interacting $\mathrm{Weyl}$ semimetals},\ }\href
	{https://doi.org/10.1103/PhysRevB.94.241102} {\bibfield  {journal} {\bibinfo
			{journal} {Phys. Rev. B}\ }\textbf {\bibinfo {volume} {94}},\ \bibinfo
		{pages} {241102} (\bibinfo {year} {2016})}\BibitemShut {NoStop}%
	\bibitem [{\citenamefont {Roy}\ \emph {et~al.}(2017)\citenamefont {Roy},
		\citenamefont {Goswami},\ and\ \citenamefont {Juri\ifmmode \check{c}\else
			\v{c}\fi{}i\ifmmode~\acute{c}\else \'{c}\fi{}}}]{PhysRevB.95.201102}%
	\BibitemOpen
	\bibfield  {author} {\bibinfo {author} {\bibfnamefont {B.}~\bibnamefont
			{Roy}}, \bibinfo {author} {\bibfnamefont {P.}~\bibnamefont {Goswami}},\ and\
		\bibinfo {author} {\bibfnamefont {V.}~\bibnamefont {Juri\ifmmode
				\check{c}\else \v{c}\fi{}i\ifmmode~\acute{c}\else \'{c}\fi{}}},\ }\bibfield
	{title} {\bibinfo {title} {Interacting $\mathrm{Weyl}$ fermions:
			$\mathrm{Phases}$, phase transitions, and global phase diagram},\ }\href
	{https://doi.org/10.1103/PhysRevB.95.201102} {\bibfield  {journal} {\bibinfo
			{journal} {Phys. Rev. B}\ }\textbf {\bibinfo {volume} {95}},\ \bibinfo
		{pages} {201102} (\bibinfo {year} {2017})}\BibitemShut {NoStop}%
	\bibitem [{\citenamefont {Liu}\ \emph {et~al.}(2022)\citenamefont {Liu},
		\citenamefont {Liu}, \citenamefont {Xu}, \citenamefont {Shen}, \citenamefont
		{Li}, \citenamefont {Pei}, \citenamefont {Liang}, \citenamefont {Dudin},
		\citenamefont {Kim}, \citenamefont {Cacho} \emph {et~al.}}]{liu2022direct}%
	\BibitemOpen
	\bibfield  {author} {\bibinfo {author} {\bibfnamefont {D.}~\bibnamefont
			{Liu}}, \bibinfo {author} {\bibfnamefont {E.}~\bibnamefont {Liu}}, \bibinfo
		{author} {\bibfnamefont {Q.}~\bibnamefont {Xu}}, \bibinfo {author}
		{\bibfnamefont {J.}~\bibnamefont {Shen}}, \bibinfo {author} {\bibfnamefont
			{Y.}~\bibnamefont {Li}}, \bibinfo {author} {\bibfnamefont {D.}~\bibnamefont
			{Pei}}, \bibinfo {author} {\bibfnamefont {A.}~\bibnamefont {Liang}}, \bibinfo
		{author} {\bibfnamefont {P.}~\bibnamefont {Dudin}}, \bibinfo {author}
		{\bibfnamefont {T.}~\bibnamefont {Kim}}, \bibinfo {author} {\bibfnamefont
			{C.}~\bibnamefont {Cacho}}, \emph {et~al.},\ }\bibfield  {title} {\bibinfo
		{title} {Direct observation of the spin-orbit coupling effect in magnetic
			$\mathrm{Weyl}$ semimetal $\mathrm{Co}_{3}\mathrm{Sn}_{2}\mathrm{S}_{2}$},\
	}\href {https://doi.org/10.1038/s41535-021-00392-9} {\bibfield  {journal}
		{\bibinfo  {journal} {npj Quantum Materials}\ }\textbf {\bibinfo {volume}
			{7}},\ \bibinfo {pages} {11} (\bibinfo {year} {2022})}\BibitemShut {NoStop}%
	\bibitem [{\citenamefont {Liu}\ \emph {et~al.}(2021)\citenamefont {Liu},
		\citenamefont {Xu}, \citenamefont {Liu}, \citenamefont {Shen}, \citenamefont
		{Le}, \citenamefont {Li}, \citenamefont {Pei}, \citenamefont {Liang},
		\citenamefont {Dudin}, \citenamefont {Kim}, \citenamefont {Cacho},
		\citenamefont {Xu}, \citenamefont {Sun}, \citenamefont {Yang}, \citenamefont
		{Liu}, \citenamefont {Felser}, \citenamefont {Parkin},\ and\ \citenamefont
		{Chen}}]{PhysRevB.104.205140}%
	\BibitemOpen
	\bibfield  {author} {\bibinfo {author} {\bibfnamefont {D.~F.}\ \bibnamefont
			{Liu}}, \bibinfo {author} {\bibfnamefont {Q.~N.}\ \bibnamefont {Xu}},
		\bibinfo {author} {\bibfnamefont {E.~K.}\ \bibnamefont {Liu}}, \bibinfo
		{author} {\bibfnamefont {J.~L.}\ \bibnamefont {Shen}}, \bibinfo {author}
		{\bibfnamefont {C.~C.}\ \bibnamefont {Le}}, \bibinfo {author} {\bibfnamefont
			{Y.~W.}\ \bibnamefont {Li}}, \bibinfo {author} {\bibfnamefont
			{D.}~\bibnamefont {Pei}}, \bibinfo {author} {\bibfnamefont {A.~J.}\
			\bibnamefont {Liang}}, \bibinfo {author} {\bibfnamefont {P.}~\bibnamefont
			{Dudin}}, \bibinfo {author} {\bibfnamefont {T.~K.}\ \bibnamefont {Kim}},
		\bibinfo {author} {\bibfnamefont {C.}~\bibnamefont {Cacho}}, \bibinfo
		{author} {\bibfnamefont {Y.~F.}\ \bibnamefont {Xu}}, \bibinfo {author}
		{\bibfnamefont {Y.}~\bibnamefont {Sun}}, \bibinfo {author} {\bibfnamefont
			{L.~X.}\ \bibnamefont {Yang}}, \bibinfo {author} {\bibfnamefont {Z.~K.}\
			\bibnamefont {Liu}}, \bibinfo {author} {\bibfnamefont {C.}~\bibnamefont
			{Felser}}, \bibinfo {author} {\bibfnamefont {S.~S.~P.}\ \bibnamefont
			{Parkin}},\ and\ \bibinfo {author} {\bibfnamefont {Y.~L.}\ \bibnamefont
			{Chen}},\ }\bibfield  {title} {\bibinfo {title} {$\mathrm{Topological}$ phase
			transition in a magnetic weyl semimetal},\ }\href
	{https://doi.org/10.1103/PhysRevB.104.205140} {\bibfield  {journal} {\bibinfo
			{journal} {Phys. Rev. B}\ }\textbf {\bibinfo {volume} {104}},\ \bibinfo
		{pages} {205140} (\bibinfo {year} {2021})}\BibitemShut {NoStop}%
	\bibitem [{\citenamefont {Yang}\ \emph
		{et~al.}(2020{\natexlab{b}})\citenamefont {Yang}, \citenamefont {Zhang},
		\citenamefont {Zhou}, \citenamefont {Dai}, \citenamefont {Liao},
		\citenamefont {Weng},\ and\ \citenamefont {Qiu}}]{PhysRevLett.124.077403}%
	\BibitemOpen
	\bibfield  {author} {\bibinfo {author} {\bibfnamefont {R.}~\bibnamefont
			{Yang}}, \bibinfo {author} {\bibfnamefont {T.}~\bibnamefont {Zhang}},
		\bibinfo {author} {\bibfnamefont {L.}~\bibnamefont {Zhou}}, \bibinfo {author}
		{\bibfnamefont {Y.}~\bibnamefont {Dai}}, \bibinfo {author} {\bibfnamefont
			{Z.}~\bibnamefont {Liao}}, \bibinfo {author} {\bibfnamefont {H.}~\bibnamefont
			{Weng}},\ and\ \bibinfo {author} {\bibfnamefont {X.}~\bibnamefont {Qiu}},\
	}\bibfield  {title} {\bibinfo {title} {Magnetization-$\mathrm{Induced}$
			$\mathrm{Band}$ $\mathrm{Shift}$ in $\mathrm{Ferromagnetic}$ $\mathrm{Weyl}$
			$\mathrm{Semimetal}$ $\mathrm{Co}_{3}\mathrm{Sn}_{2}\mathrm{S}_{2}$},\ }\href
	{https://doi.org/10.1103/PhysRevLett.124.077403} {\bibfield  {journal}
		{\bibinfo  {journal} {Phys. Rev. Lett.}\ }\textbf {\bibinfo {volume} {124}},\
		\bibinfo {pages} {077403} (\bibinfo {year} {2020}{\natexlab{b}})}\BibitemShut
	{NoStop}%
	\bibitem [{SI()}]{SI}%
	\BibitemOpen
	\bibfield  {title} {\bibinfo {title} {Support},\ }\href@noop {} {\bibinfo
		{journal} {See Supplemental Material for more imformation.}\ }\BibitemShut
	{NoStop}%
	\bibitem [{\citenamefont {Fu}\ and\ \citenamefont
		{Kane}(2007)}]{PhysRevB.76.045302}%
	\BibitemOpen
	\bibfield  {journal} {  }\bibfield  {author} {\bibinfo {author} {\bibfnamefont
			{L.}~\bibnamefont {Fu}}\ and\ \bibinfo {author} {\bibfnamefont {C.~L.}\
			\bibnamefont {Kane}},\ }\bibfield  {title} {\bibinfo {title}
		{$\mathrm{Topological}$ insulators with inversion symmetry},\ }\href
	{https://doi.org/10.1103/PhysRevB.76.045302} {\bibfield  {journal} {\bibinfo
			{journal} {Phys. Rev. B}\ }\textbf {\bibinfo {volume} {76}},\ \bibinfo
		{pages} {045302} (\bibinfo {year} {2007})}\BibitemShut {NoStop}%
	\bibitem [{\citenamefont {Mathai}\ and\ \citenamefont
		{Thiang}(2017)}]{Mathai_2017}%
	\BibitemOpen
	\bibfield  {author} {\bibinfo {author} {\bibfnamefont {V.}~\bibnamefont
			{Mathai}}\ and\ \bibinfo {author} {\bibfnamefont {G.~C.}\ \bibnamefont
			{Thiang}},\ }\bibfield  {title} {\bibinfo {title} {Global topology of
			$\mathrm{Weyl}$ semimetals and $\mathrm{Fermi}$ arcs},\ }\href
	{https://doi.org/10.1088/1751-8121/aa59b2} {\bibfield  {journal} {\bibinfo
			{journal} {Journal of Physics A: Mathematical and Theoretical}\ }\textbf
		{\bibinfo {volume} {50}},\ \bibinfo {pages} {11LT01} (\bibinfo {year}
		{2017})}\BibitemShut {NoStop}%
	\bibitem [{\citenamefont {Jia}\ \emph {et~al.}(2016)\citenamefont {Jia},
		\citenamefont {Xu},\ and\ \citenamefont {Hasan}}]{jia2016weyl}%
	\BibitemOpen
	\bibfield  {author} {\bibinfo {author} {\bibfnamefont {S.}~\bibnamefont
			{Jia}}, \bibinfo {author} {\bibfnamefont {S.-Y.}\ \bibnamefont {Xu}},\ and\
		\bibinfo {author} {\bibfnamefont {M.~Z.}\ \bibnamefont {Hasan}},\ }\bibfield
	{title} {\bibinfo {title} {Weyl semimetals, $\mathrm{Fermi}$ arcs and chiral
			anomalies},\ }\href {https://doi.org/10.1038/nmat4787} {\bibfield  {journal}
		{\bibinfo  {journal} {Nature Mterials}\ }\textbf {\bibinfo {volume} {15}},\
		\bibinfo {pages} {1140} (\bibinfo {year} {2016})}\BibitemShut {NoStop}%
	\bibitem [{\citenamefont {Fang}\ \emph {et~al.}(2016)\citenamefont {Fang},
		\citenamefont {Lu},\ and\ \citenamefont {Fu}}]{nphys3782}%
	\BibitemOpen
	\bibfield  {author} {\bibinfo {author} {\bibfnamefont {C.}~\bibnamefont
			{Fang}}, \bibinfo {author} {\bibfnamefont {L.}~\bibnamefont {Lu}},\ and\
		\bibinfo {author} {\bibfnamefont {L.}~\bibnamefont {Fu}},\ }\bibfield
	{title} {\bibinfo {title} {Topological semimetals with helicoid
			surface states},\ }\href {https://doi.org/10.1038/nphys3782} {\bibfield
		{journal} {\bibinfo  {journal} {Nature Physics}\ }\textbf {\bibinfo {volume}
			{12}},\ \bibinfo {pages} {936} (\bibinfo {year} {2016})}\BibitemShut
	{NoStop}%
	\bibitem [{\citenamefont {Wang}\ \emph {et~al.}(2017)\citenamefont {Wang},
		\citenamefont {Lin}, \citenamefont {Wang}, \citenamefont {Yu},\ and\
		\citenamefont {and}}]{Wang04052017}%
	\BibitemOpen
	\bibfield  {author} {\bibinfo {author} {\bibfnamefont {S.}~\bibnamefont
			{Wang}}, \bibinfo {author} {\bibfnamefont {B.-C.}\ \bibnamefont {Lin}},
		\bibinfo {author} {\bibfnamefont {A.-Q.}\ \bibnamefont {Wang}}, \bibinfo
		{author} {\bibfnamefont {D.-P.}\ \bibnamefont {Yu}},\ and\ \bibinfo {author}
		{\bibfnamefont {Z.-M.~L.}\ \bibnamefont {and}},\ }\bibfield  {title}
	{\bibinfo {title} {Quantum transport in dirac and weyl semimetals: a
			review},\ }\href {https://doi.org/10.1080/23746149.2017.1327329} {\bibfield
		{journal} {\bibinfo  {journal} {Advances in Physics: X}\ }\textbf {\bibinfo
			{volume} {2}},\ \bibinfo {pages} {518} (\bibinfo {year} {2017})}\BibitemShut
	{NoStop}%
	\bibitem [{\citenamefont {Yang}\ \emph {et~al.}(2019)\citenamefont {Yang},
		\citenamefont {Yang}, \citenamefont {Liu}, \citenamefont {Sun}, \citenamefont
		{Chen}, \citenamefont {Peng}, \citenamefont {Schmidt}, \citenamefont
		{Prabhakaran}, \citenamefont {Bernevig}, \citenamefont {Felser} \emph
		{et~al.}}]{yang2019topological}%
	\BibitemOpen
	\bibfield  {author} {\bibinfo {author} {\bibfnamefont {H.}~\bibnamefont
			{Yang}}, \bibinfo {author} {\bibfnamefont {L.}~\bibnamefont {Yang}}, \bibinfo
		{author} {\bibfnamefont {Z.}~\bibnamefont {Liu}}, \bibinfo {author}
		{\bibfnamefont {Y.}~\bibnamefont {Sun}}, \bibinfo {author} {\bibfnamefont
			{C.}~\bibnamefont {Chen}}, \bibinfo {author} {\bibfnamefont {H.}~\bibnamefont
			{Peng}}, \bibinfo {author} {\bibfnamefont {M.}~\bibnamefont {Schmidt}},
		\bibinfo {author} {\bibfnamefont {D.}~\bibnamefont {Prabhakaran}}, \bibinfo
		{author} {\bibfnamefont {B.~A.}\ \bibnamefont {Bernevig}}, \bibinfo {author}
		{\bibfnamefont {C.}~\bibnamefont {Felser}}, \emph {et~al.},\ }\bibfield
	{title} {\bibinfo {title} {Topological $\mathrm{Lifshitz}$ transitions and
			$\mathrm{Fermi}$ arc manipulation in $\mathrm{Weyl}$ semimetal
			$\mathrm{NbAs}$},\ }\href {https://doi.org/10.1038/s41467-019-11491-4}
	{\bibfield  {journal} {\bibinfo  {journal} {Nature Communications}\ }\textbf
		{\bibinfo {volume} {10}},\ \bibinfo {pages} {3478} (\bibinfo {year}
		{2019})}\BibitemShut {NoStop}%
	\bibitem [{\citenamefont {Inoue}\ \emph {et~al.}(2016)\citenamefont {Inoue},
		\citenamefont {Gyenis}, \citenamefont {Wang}, \citenamefont {Li},
		\citenamefont {Oh}, \citenamefont {Jiang}, \citenamefont {Ni}, \citenamefont
		{Bernevig},\ and\ \citenamefont {Yazdani}}]{science.aad8766}%
	\BibitemOpen
	\bibfield  {author} {\bibinfo {author} {\bibfnamefont {H.}~\bibnamefont
			{Inoue}}, \bibinfo {author} {\bibfnamefont {A.}~\bibnamefont {Gyenis}},
		\bibinfo {author} {\bibfnamefont {Z.}~\bibnamefont {Wang}}, \bibinfo {author}
		{\bibfnamefont {J.}~\bibnamefont {Li}}, \bibinfo {author} {\bibfnamefont
			{S.~W.}\ \bibnamefont {Oh}}, \bibinfo {author} {\bibfnamefont
			{S.}~\bibnamefont {Jiang}}, \bibinfo {author} {\bibfnamefont
			{N.}~\bibnamefont {Ni}}, \bibinfo {author} {\bibfnamefont {B.~A.}\
			\bibnamefont {Bernevig}},\ and\ \bibinfo {author} {\bibfnamefont
			{A.}~\bibnamefont {Yazdani}},\ }\bibfield  {title} {\bibinfo {title}
		{Quasiparticle interference of the $\mathrm{Fermi}$ arcs and surface-bulk
			connectivity of a $\mathrm{Weyl}$ semimetal},\ }\href
	{https://doi.org/10.1126/science.aad8766} {\bibfield  {journal} {\bibinfo
			{journal} {Science}\ }\textbf {\bibinfo {volume} {351}},\ \bibinfo {pages}
		{1184} (\bibinfo {year} {2016})}\BibitemShut {NoStop}%
	\bibitem [{\citenamefont {Duan}\ \emph {et~al.}(2023)\citenamefont {Duan},
		\citenamefont {Lu},\ and\ \citenamefont {Liu}}]{PhysRevB.108.195436}%
	\BibitemOpen
	\bibfield  {author} {\bibinfo {author} {\bibfnamefont {W.}~\bibnamefont
			{Duan}}, \bibinfo {author} {\bibfnamefont {X.}~\bibnamefont {Lu}},\ and\
		\bibinfo {author} {\bibfnamefont {J.-F.}\ \bibnamefont {Liu}},\ }\bibfield
	{title} {\bibinfo {title} {Large optical conductivity of $\mathrm{Fermi}$ arc
			states in $\mathrm{Weyl}$ and $\mathrm{Dirac}$ semimetal nanowires},\ }\href
	{https://doi.org/10.1103/PhysRevB.108.195436} {\bibfield  {journal} {\bibinfo
			{journal} {Phys. Rev. B}\ }\textbf {\bibinfo {volume} {108}},\ \bibinfo
		{pages} {195436} (\bibinfo {year} {2023})}\BibitemShut {NoStop}%
	\bibitem [{\citenamefont {Hosur}(2012)}]{PhysRevB.86.195102}%
	\BibitemOpen
	\bibfield  {author} {\bibinfo {author} {\bibfnamefont {P.}~\bibnamefont
			{Hosur}},\ }\bibfield  {title} {\bibinfo {title} {Friedel oscillations due to
			$\mathrm{Fermi}$ arcs in $\mathrm{Weyl}$ semimetals},\ }\href
	{https://doi.org/10.1103/PhysRevB.86.195102} {\bibfield  {journal} {\bibinfo
			{journal} {Phys. Rev. B}\ }\textbf {\bibinfo {volume} {86}},\ \bibinfo
		{pages} {195102} (\bibinfo {year} {2012})}\BibitemShut {NoStop}%
	\bibitem [{\citenamefont {Breitkreiz}\ and\ \citenamefont
		{Brouwer}(2023)}]{PhysRevLett.130.196602}%
	\BibitemOpen
	\bibfield  {author} {\bibinfo {author} {\bibfnamefont {M.}~\bibnamefont
			{Breitkreiz}}\ and\ \bibinfo {author} {\bibfnamefont {P.~W.}\ \bibnamefont
			{Brouwer}},\ }\bibfield  {title} {\bibinfo {title} {Fermi-$\mathrm{Arc}$
			$\mathrm{Metals}$},\ }\href {https://doi.org/10.1103/PhysRevLett.130.196602}
	{\bibfield  {journal} {\bibinfo  {journal} {Phys. Rev. Lett.}\ }\textbf
		{\bibinfo {volume} {130}},\ \bibinfo {pages} {196602} (\bibinfo {year}
		{2023})}\BibitemShut {NoStop}%
	\bibitem [{\citenamefont {Sukhachov}\ \emph {et~al.}(2020)\citenamefont
		{Sukhachov}, \citenamefont {Rakov}, \citenamefont {Teslyk},\ and\
		\citenamefont {Gorbar}}]{andp.201900449}%
	\BibitemOpen
	\bibfield  {author} {\bibinfo {author} {\bibfnamefont {P.~O.}\ \bibnamefont
			{Sukhachov}}, \bibinfo {author} {\bibfnamefont {M.~V.}\ \bibnamefont
			{Rakov}}, \bibinfo {author} {\bibfnamefont {O.~M.}\ \bibnamefont {Teslyk}},\
		and\ \bibinfo {author} {\bibfnamefont {E.~V.}\ \bibnamefont {Gorbar}},\
	}\bibfield  {title} {\bibinfo {title} {Fermi $\mathrm{Arcs}$ and
			$\mathrm{DC}$ $\mathrm{Transport}$ in $\mathrm{Nanowires}$ of
			$\mathrm{Dirac}$ and $\mathrm{Weyl}$ $\mathrm{Semimetals}$},\ }\href
	{https://doi.org/https://doi.org/10.1002/andp.201900449} {\bibfield
		{journal} {\bibinfo  {journal} {Annalen der Physik}\ }\textbf {\bibinfo
			{volume} {532}},\ \bibinfo {pages} {1900449} (\bibinfo {year}
		{2020})}\BibitemShut {NoStop}%
	\bibitem [{\citenamefont {Blaha}\ \emph {et~al.}(2020)\citenamefont {Blaha},
		\citenamefont {Schwarz}, \citenamefont {Tran}, \citenamefont {Laskowski},
		\citenamefont {Madsen},\ and\ \citenamefont {Marks}}]{10.1063/1.5143061}%
	\BibitemOpen
	\bibfield  {author} {\bibinfo {author} {\bibfnamefont {P.}~\bibnamefont
			{Blaha}}, \bibinfo {author} {\bibfnamefont {K.}~\bibnamefont {Schwarz}},
		\bibinfo {author} {\bibfnamefont {F.}~\bibnamefont {Tran}}, \bibinfo {author}
		{\bibfnamefont {R.}~\bibnamefont {Laskowski}}, \bibinfo {author}
		{\bibfnamefont {G.~K.~H.}\ \bibnamefont {Madsen}},\ and\ \bibinfo {author}
		{\bibfnamefont {L.~D.}\ \bibnamefont {Marks}},\ }\bibfield  {title} {\bibinfo
		{title} {Wien2k: $\mathrm{An}$ $\mathrm{APW+lo}$ $\mathrm{Program}$ for
			$\mathrm{Calculating}$ the $\mathrm{Properties}$ of $\mathrm{Solids}$},\
	}\href {https://doi.org/10.1063/1.5143061} {\bibfield  {journal} {\bibinfo
			{journal} {The Journal of Chemical Physics}\ }\textbf {\bibinfo {volume}
			{152}},\ \bibinfo {pages} {074101} (\bibinfo {year} {2020})}\BibitemShut
	{NoStop}%
	\bibitem [{\citenamefont {Perdew}\ \emph {et~al.}(1996)\citenamefont {Perdew},
		\citenamefont {Burke},\ and\ \citenamefont
		{Ernzerhof}}]{PhysRevLett.77.3865}%
	\BibitemOpen
	\bibfield  {author} {\bibinfo {author} {\bibfnamefont {J.~P.}\ \bibnamefont
			{Perdew}}, \bibinfo {author} {\bibfnamefont {K.}~\bibnamefont {Burke}},\ and\
		\bibinfo {author} {\bibfnamefont {M.}~\bibnamefont {Ernzerhof}},\ }\bibfield
	{title} {\bibinfo {title} {Generalized $\mathrm{Gradient}$
			$\mathrm{Approximation}$ $\mathrm{Made}$ $\mathrm{Simple}$},\ }\href
	{https://doi.org/10.1103/PhysRevLett.77.3865} {\bibfield  {journal} {\bibinfo
			{journal} {Phys. Rev. Lett.}\ }\textbf {\bibinfo {volume} {77}},\ \bibinfo
		{pages} {3865} (\bibinfo {year} {1996})}\BibitemShut {NoStop}%
	\bibitem [{\citenamefont {Kresse}\ and\ \citenamefont
		{Furthmüller}(1996)}]{KRESSE199615}%
	\BibitemOpen
	\bibfield  {author} {\bibinfo {author} {\bibfnamefont {G.}~\bibnamefont
			{Kresse}}\ and\ \bibinfo {author} {\bibfnamefont {J.}~\bibnamefont
			{Furthmüller}},\ }\bibfield  {title} {\bibinfo {title} {Efficiency of
			ab-initio total energy calculations for metals and semiconductors using a
			plane-wave basis set},\ }\href
	{https://doi.org/https://doi.org/10.1016/0927-0256(96)00008-0} {\bibfield
		{journal} {\bibinfo  {journal} {Computational Materials Science}\ }\textbf
		{\bibinfo {volume} {6}},\ \bibinfo {pages} {15} (\bibinfo {year}
		{1996})}\BibitemShut {NoStop}%
	\bibitem [{\citenamefont {Kresse}\ and\ \citenamefont
		{Furthm\"uller}(1996)}]{PhysRevB.54.11169}%
	\BibitemOpen
	\bibfield  {author} {\bibinfo {author} {\bibfnamefont {G.}~\bibnamefont
			{Kresse}}\ and\ \bibinfo {author} {\bibfnamefont {J.}~\bibnamefont
			{Furthm\"uller}},\ }\bibfield  {title} {\bibinfo {title} {Efficient iterative
			schemes for ab initio total-energy calculations using a plane-wave basis
			set},\ }\href {https://doi.org/10.1103/PhysRevB.54.11169} {\bibfield
		{journal} {\bibinfo  {journal} {Phys. Rev. B}\ }\textbf {\bibinfo {volume}
			{54}},\ \bibinfo {pages} {11169} (\bibinfo {year} {1996})}\BibitemShut
	{NoStop}%
	\bibitem [{\citenamefont {Haule}(2007)}]{PhysRevB.75.155113}%
	\BibitemOpen
	\bibfield  {author} {\bibinfo {author} {\bibfnamefont {K.}~\bibnamefont
			{Haule}},\ }\bibfield  {title} {\bibinfo {title} {Quantum $\mathrm{Monte}$
			$\mathrm{Carlo}$ impurity solver for cluster dynamical mean-field theory and
			electronic structure calculations with adjustable cluster base},\ }\href
	{https://doi.org/10.1103/PhysRevB.75.155113} {\bibfield  {journal} {\bibinfo
			{journal} {Phys. Rev. B}\ }\textbf {\bibinfo {volume} {75}},\ \bibinfo
		{pages} {155113} (\bibinfo {year} {2007})}\BibitemShut {NoStop}%
	\bibitem [{\citenamefont {Werner}\ \emph {et~al.}(2006)\citenamefont {Werner},
		\citenamefont {Comanac}, \citenamefont {de' Medici}, \citenamefont {Troyer},\
		and\ \citenamefont {Millis}}]{PhysRevLett.97.076405}%
	\BibitemOpen
	\bibfield  {author} {\bibinfo {author} {\bibfnamefont {P.}~\bibnamefont
			{Werner}}, \bibinfo {author} {\bibfnamefont {A.}~\bibnamefont {Comanac}},
		\bibinfo {author} {\bibfnamefont {L.}~\bibnamefont {de' Medici}}, \bibinfo
		{author} {\bibfnamefont {M.}~\bibnamefont {Troyer}},\ and\ \bibinfo {author}
		{\bibfnamefont {A.~J.}\ \bibnamefont {Millis}},\ }\bibfield  {title}
	{\bibinfo {title} {Continuous-$\mathrm{Time}$ $\mathrm{Solver}$ for
			$\mathrm{Quantum}$ $\mathrm{Impurity}$ $\mathrm{Models}$},\ }\href
	{https://doi.org/10.1103/PhysRevLett.97.076405} {\bibfield  {journal}
		{\bibinfo  {journal} {Phys. Rev. Lett.}\ }\textbf {\bibinfo {volume} {97}},\
		\bibinfo {pages} {076405} (\bibinfo {year} {2006})}\BibitemShut {NoStop}%
	\bibitem [{\citenamefont {Haule}\ \emph {et~al.}(2010)\citenamefont {Haule},
		\citenamefont {Yee},\ and\ \citenamefont {Kim}}]{PhysRevB.81.195107}%
	\BibitemOpen
	\bibfield  {author} {\bibinfo {author} {\bibfnamefont {K.}~\bibnamefont
			{Haule}}, \bibinfo {author} {\bibfnamefont {C.-H.}\ \bibnamefont {Yee}},\
		and\ \bibinfo {author} {\bibfnamefont {K.}~\bibnamefont {Kim}},\ }\bibfield
	{title} {\bibinfo {title} {Dynamical mean-field theory within the
			full-potential methods: Electronic structure of $\mathrm{{CeIrIn}_{5}}$,
			$\mathrm{{CeCoIn}_{5}}$, and $\mathrm{{CeRhIn}_{5}}$},\ }\href
	{https://doi.org/10.1103/PhysRevB.81.195107} {\bibfield  {journal} {\bibinfo
			{journal} {Phys. Rev. B}\ }\textbf {\bibinfo {volume} {81}},\ \bibinfo
		{pages} {195107} (\bibinfo {year} {2010})}\BibitemShut {NoStop}%
	\bibitem [{\citenamefont {Marzari}\ and\ \citenamefont
		{Vanderbilt}(1997)}]{PhysRevB.56.12847}%
	\BibitemOpen
	\bibfield  {author} {\bibinfo {author} {\bibfnamefont {N.}~\bibnamefont
			{Marzari}}\ and\ \bibinfo {author} {\bibfnamefont {D.}~\bibnamefont
			{Vanderbilt}},\ }\bibfield  {title} {\bibinfo {title} {Maximally localized
			generalized $\mathrm{Wannier}$ functions for composite energy bands},\ }\href
	{https://doi.org/10.1103/PhysRevB.56.12847} {\bibfield  {journal} {\bibinfo
			{journal} {Phys. Rev. B}\ }\textbf {\bibinfo {volume} {56}},\ \bibinfo
		{pages} {12847} (\bibinfo {year} {1997})}\BibitemShut {NoStop}%
	\bibitem [{\citenamefont {Souza}\ \emph {et~al.}(2001)\citenamefont {Souza},
		\citenamefont {Marzari},\ and\ \citenamefont
		{Vanderbilt}}]{PhysRevB.65.035109}%
	\BibitemOpen
	\bibfield  {author} {\bibinfo {author} {\bibfnamefont {I.}~\bibnamefont
			{Souza}}, \bibinfo {author} {\bibfnamefont {N.}~\bibnamefont {Marzari}},\
		and\ \bibinfo {author} {\bibfnamefont {D.}~\bibnamefont {Vanderbilt}},\
	}\bibfield  {title} {\bibinfo {title} {Maximally localized $\mathrm{Wannier}$
			functions for entangled energy bands},\ }\href
	{https://doi.org/10.1103/PhysRevB.65.035109} {\bibfield  {journal} {\bibinfo
			{journal} {Phys. Rev. B}\ }\textbf {\bibinfo {volume} {65}},\ \bibinfo
		{pages} {035109} (\bibinfo {year} {2001})}\BibitemShut {NoStop}%
	\bibitem [{\citenamefont {Marzari}\ \emph {et~al.}(2012)\citenamefont
		{Marzari}, \citenamefont {Mostofi}, \citenamefont {Yates}, \citenamefont
		{Souza},\ and\ \citenamefont {Vanderbilt}}]{RevModPhys.84.1419}%
	\BibitemOpen
	\bibfield  {author} {\bibinfo {author} {\bibfnamefont {N.}~\bibnamefont
			{Marzari}}, \bibinfo {author} {\bibfnamefont {A.~A.}\ \bibnamefont
			{Mostofi}}, \bibinfo {author} {\bibfnamefont {J.~R.}\ \bibnamefont {Yates}},
		\bibinfo {author} {\bibfnamefont {I.}~\bibnamefont {Souza}},\ and\ \bibinfo
		{author} {\bibfnamefont {D.}~\bibnamefont {Vanderbilt}},\ }\bibfield  {title}
	{\bibinfo {title} {Maximally localized $\mathrm{Wannier}$ functions:
			$\mathrm{Theory}$ and applications},\ }\href
	{https://doi.org/10.1103/RevModPhys.84.1419} {\bibfield  {journal} {\bibinfo
			{journal} {Rev. Mod. Phys.}\ }\textbf {\bibinfo {volume} {84}},\ \bibinfo
		{pages} {1419} (\bibinfo {year} {2012})}\BibitemShut {NoStop}%
	\bibitem [{\citenamefont {Mostofi}\ \emph {et~al.}(2014)\citenamefont
		{Mostofi}, \citenamefont {Yates}, \citenamefont {Pizzi}, \citenamefont {Lee},
		\citenamefont {Souza}, \citenamefont {Vanderbilt},\ and\ \citenamefont
		{Marzari}}]{MOSTOFI20142309}%
	\BibitemOpen
	\bibfield  {author} {\bibinfo {author} {\bibfnamefont {A.~A.}\ \bibnamefont
			{Mostofi}}, \bibinfo {author} {\bibfnamefont {J.~R.}\ \bibnamefont {Yates}},
		\bibinfo {author} {\bibfnamefont {G.}~\bibnamefont {Pizzi}}, \bibinfo
		{author} {\bibfnamefont {Y.-S.}\ \bibnamefont {Lee}}, \bibinfo {author}
		{\bibfnamefont {I.}~\bibnamefont {Souza}}, \bibinfo {author} {\bibfnamefont
			{D.}~\bibnamefont {Vanderbilt}},\ and\ \bibinfo {author} {\bibfnamefont
			{N.}~\bibnamefont {Marzari}},\ }\bibfield  {title} {\bibinfo {title} {An
			updated version of wannier90: A tool for obtaining maximally-localised
			$\mathrm{Wannier}$ functions},\ }\href
	{https://doi.org/https://doi.org/10.1016/j.cpc.2014.05.003} {\bibfield
		{journal} {\bibinfo  {journal} {Computer Physics Communications}\ }\textbf
		{\bibinfo {volume} {185}},\ \bibinfo {pages} {2309} (\bibinfo {year}
		{2014})}\BibitemShut {NoStop}%
	\bibitem [{\citenamefont {Wu}\ \emph {et~al.}(2018)\citenamefont {Wu},
		\citenamefont {Zhang}, \citenamefont {Song}, \citenamefont {Troyer},\ and\
		\citenamefont {Soluyanov}}]{WU2018405}%
	\BibitemOpen
	\bibfield  {author} {\bibinfo {author} {\bibfnamefont {Q.}~\bibnamefont
			{Wu}}, \bibinfo {author} {\bibfnamefont {S.}~\bibnamefont {Zhang}}, \bibinfo
		{author} {\bibfnamefont {H.-F.}\ \bibnamefont {Song}}, \bibinfo {author}
		{\bibfnamefont {M.}~\bibnamefont {Troyer}},\ and\ \bibinfo {author}
		{\bibfnamefont {A.~A.}\ \bibnamefont {Soluyanov}},\ }\bibfield  {title}
	{\bibinfo {title} {Wanniertools: $\mathrm{An}$ open-source software package
			for novel topological materials},\ }\href
	{https://doi.org/https://doi.org/10.1016/j.cpc.2017.09.033} {\bibfield
		{journal} {\bibinfo  {journal} {Computer Physics Communications}\ }\textbf
		{\bibinfo {volume} {224}},\ \bibinfo {pages} {405} (\bibinfo {year}
		{2018})}\BibitemShut {NoStop}%
\end{thebibliography}
\end{document}